\documentclass[fleqn,usenatbib]{mnras}
\usepackage[T1]{fontenc}
\usepackage{ae,aecompl}
\usepackage{relsize} 
\usepackage{color,soul}
\usepackage{subfig}
\usepackage{graphicx}
\usepackage{amsmath}
\usepackage{amssymb}
\usepackage{bm}
\usepackage{footnote}
\usepackage{bigints}
\usepackage{colortbl}
\defcitealias{Gie11}{GH11}
\defcitealias{Bon03}{BCB+03}
\defcitealias{Gil00}{GBG+00}
\definecolor{Gray}{gray}{0.9}

\usepackage{newtxtext,newtxmath}

\title[Brown dwarf tidal capture in globular clusters]{\vspace{-2mm}Forming short period sub-stellar companions in 47 Tuc: I.~Dynamical model and brown dwarf tidal capture rates\vspace{-3mm}}
\author[A.~J.~Winter et al.]{Andrew~J.~Winter,$^{1,2,3}$\thanks{andrew.winter@uni-heidelberg.de} Giovanni P. Rosotti$^{2,3,4}$, Cathie Clarke$^{2}$, Mirek Giersz$^{5}$ 	\\
$^{1}$Zentrum f\"{u}r Astronomie, Heidelberg University, Albert Ueberle Str. 2, 69120 Heidelberg, Germany \\
$^{2}$Institute of Astronomy, University of Cambridge, Madingley Road, Cambridge CB3 0HA, UK \\
$^{3}$School of Physics and Astronomy, University of Leicester, Leicester, LE1 7RH, UK \\
$^{4}$Leiden Observatory, Leiden University, P.O. Box 9513, NL-2300 RA Leiden, the Netherlands\\
$^{5}$Nicolaus Copernicus Astronomical Center, Polish Academy of Sciences, ul. Bartycka 18, Warsaw 00-716 Poland
\vspace{-3mm}
}
\date{Accepted X{\sevensize xxxx} XX. Received X{\sevensize xxxx} XX; in original form 2019 April XX}\vspace{-2mm}
\pubyear{2021}
\begin{document}
\label{firstpage}
\pagerange{\pageref{firstpage}--\pageref{lastpage}}
\maketitle
\begin{abstract}
Stars in globular clusters formed and evolved in the most extreme environment: high density and low metallicity. If the formation of stars and planets are at all sensitive to environmental conditions, this should therefore be evident in globular clusters. Observations have indicated that hot Jupiters are at least an order of magnitude less prevalent in the central region of the globular cluster 47 Tucanae than in the field. In this work, we explore the claims in the literature for additional consequences for the low mass stellar initial mass function. Tidal capture, the mechanism that produces X-ray binaries in globular clusters, applies also to brown dwarfs (BDs). This process produces tight stellar-BD binaries that would be detectable by transit surveys. Applying a Monte Carlo dynamical evolution model, we compute the overall BD capture rates. We find that the number of captures is lower than previous estimates. Capture efficiency increases steeply with stellar mass, which means that mass segregation reduces capture efficiency as BDs and low mass stars occupy the same regions. The result of this effect is that the current constraints on the short period companion fraction remains marginally consistent with initially equal numbers of BDs and stars. However, our findings suggest that expanding the sample in 47 Tuc or surveying other globular clusters for close sub-stellar companions can yield constraints on the sub-stellar initial mass function in these environments. {We estimate the capture rates in other globular clusters and {suggest that 47 Tuc remains a promising target for future transit surveys}.} 
\end{abstract}
\begin{keywords} 
stars:  brown dwarfs, formation, kinematics and dynamics -- globular clusters: individual: 47 Tucanae \vspace{-2mm}
\end{keywords}


\section{Introduction}

One of the most important observable predictions for star formation theory is the resulting spectrum of masses, known as the initial mass function (IMF). The extreme ends of this distribution are of particular interest for constraining star formation physics. At the lowest masses, brown dwarfs (BDs) are thought to be approximately as numerous in the Milky Way as hydrogen-burning stars \citep{Cha03}. However, what determines the IMF, and whether it varies with environment, remain the topic of debate \citep[see][for a review]{Kru14}.

BDs are usually defined to be objects less massive than required to burn ordinary hydrogen ($\lesssim 0.08\, M_\odot$) and greater than the deuterium-burning limit ($\gtrsim 13\, M_\mathrm{J}$). If the IMF in this regime is set by the local thermal Jeans mass, then the lower mass limit for fragmentation is set by the requirement that the centre of the clump can cool \citep[i.e. the opacity limit --][]{Low76}. In this case, as the metallicity increases, the cooling becomes more efficient and the minimum mass for BDs decreases \citep{Bat05}. However, dynamical processes and competitive accretion might further influence the distribution of stellar masses \citep[e.g.][]{Bon97, Kle98,Bon08, Bat12}. Dynamical interactions can eject stars from their formation environment, shutting off accretion and stunting growth before the hydrogen mass burning limit is reached \citep{Rei01, Bat02b}. Alternatively, the photoevaporation of the gas in an accreting envelope due to irradiation by neighbouring OB stars can have a similar effect \citep{Hes96, Whi04}.

The low-mass IMF appears to have a similar shape across numerous local star forming regions \citep{Andersen08}. However, a number of regions also appear to exhibit deviations from a `universal' IMF. \citet{Scholz13} and \citet{Luhman16} find that the cluster NGC 1333 has a greater fraction of sub-stellar objects than IC348, which is interpreted as evidence of enhanced low mass star formation in dense environments. For the most massive and dense local star forming region, the Orion Nebula cluster (ONC), contradictory results have been inferred by different authors. \citet{DaRio12} find a deficiency of BDs, in direct contrast to a previously inferred enhancement \citep{Muench02}. Such results might be reconciled by a {bimodal} distribution, as found by \citet{Drass16}. In this case, the enhancement in occurrence rates at the lowest BD masses may be the result of a distinct formation mechanism, possibly within a protoplanetary disc. 

The distinction between BDs and planets, while motivated by a physical threshold, may not cleanly delineate formation mechanisms \citep{Chabrier14}. In particular, planets with masses greater than $\sim 4\, M_\mathrm{J}$ may form by gravitational instability rather than core accretion \citep{Schlaufman18}. Interestingly, the masses of `planets' thought to be responsible for the gaps observed in the dust emission in protoplanetary discs with ALMA \citep[e.g.][]{HLTau_15, Muller18} can be several Jupiter masses, approaching or exceeding the canonical minimum BD mass \citep[e.g.][]{Haffert19, Christaens19}. The questions of BD and massive planet occurrence rates may therefore be related. 

Globular clusters are massive populations of stars that remain bound against galactic tides despite their old ages. Having low metallicities and high densities, they represent the present day remnant of formation in a completely different environment to local star forming regions. As such, if the formation of stars or planets is in any way dependent on environment, one should expect to see differences in globular cluster populations with respect to those in the field. \citet[][hereafter \citetalias{Gil00}]{Gil00} presented a HST survey of the globular cluster 47 Tuc in search of transiting giant planets. The null result put upper-limits on the frequency of hot Jupiters (with orbits $\lesssim 10$~days), suggesting a frequency at least an order of magnitude lower than the solar neighbourhood average \citep[$\sim 1$~percent --][]{Wright12}.

The potential significance of the dearth of low mass companions of stars in 47 Tuc compounds when one considers the potential for tidal capture in such a dense stellar environment. Tidal capture for close binary formation in globular clusters was suggested by \citet{Fab75} as a means to explain their high observed X-ray luminosity-to-mass ratios. {This two-body capture mechanism has since been incorporated into Monte Carlo simulations of globular clusters alongside three-body gravitational interactions between `point masses' \citep{Stodolkiewicz85, Stodolkiewicz86}.} \citet[][hereafter \citetalias{Bon03}]{Bon03} pointed out that the same principle also applies to the capture of BDs. This means that if the fraction of BDs that formed in globular clusters is similar to the galactic field, then a significant number of close BD binaries should also exist. Thus, the absence of detected transits in 47 Tuc not only suggests a reduced occurrence rate of hot Jupiters, but also a dearth of BDs. 

{In this mini-series of two papers, we explore the dearth of short period sub-stellar companions in 47 Tuc in terms of the expected rates of both BDs (this paper) and hot Jupiters (a second paper). To this end we apply the \textsc{Mocca} Monte Carlo code \citep{Hyp13, Giersz13} to accurately compute the theoretical capture and scattering rates over the lifetime of 47 Tuc. In this, the first of the two papers, we focus on the possibility of tidal BD capture in 47 Tuc while a second paper w ill deal with the migration of massive planets to short period orbits. We first discuss the theory of tidal capture in Section~\ref{sec:tidal_capt_calcs}. We then introduce our Monte Carlo model for 47 Tuc in Section~\ref{sec:Num_Method}. In Section~\ref{sec:results} we present the resultant BD capture rates over the lifetime of 47 Tuc, compare to the observational constraints and discuss future observations and generalisation to other globular clusters. Conclusions are summarised in Section~\ref{sec:conclusions}.  }

\section{Tidal capture theory}
\label{sec:tidal_capt_calcs}
\subsection{Tidal capture cross section}
\label{sec:tid_cross}
The tidal capture condition for a star and BD pair is discussed in detail by \citetalias{Bon03}; we briefly review the relevant equations here. During the close passage between stars, the orbital energy can be dissipated by non-radial oscillations within the stellar interiors \citep{Rob68}. It follows that if the passage is sufficiently close, the tidal dissipation can result in capture and the formation of a tight binary. \citet{Fab75} applied this mechanism to explain low mass X-ray binaries found in globular clusters. We consider a primary star of radius $R_*$, mass $m_*$. Then for a periastron distance $a_\mathrm{p}$ during an encounter with a much smaller (point) mass $m$, the condition for capture can be approximated \citep{Fab75}:
\begin{equation}
\label{eq:capt_rad}
 \frac{a_\mathrm{p}}{R_*} < \frac{a_\mathrm{capt}}{R_*} = \left[ \frac{Gm_*}{R_* v_\infty^2} q (1+q) \right]^{1/6}
\end{equation} where $q=m/m_*$, for secondary mass $m$, and $v_\infty$ is the relative velocities of the two stars at infinite separation, $a_\mathrm{capt}$ is the capture radius. The point mass approximation is justified since tides can only be excited in a much smaller secondary if the impact parameter is within the collisional cross section. Equation~\ref{eq:capt_rad} remains valid for low mass main-sequence primaries \citep[$n=3/2$ polytropes,][]{Lee86}. 

The capture radius must also exceed the sum of the radii of the two interacting bodies, otherwise the objects would collide (i.e. periastron distance $a_\mathrm{p}> R_*+R_\mathrm{bd}$, where $R_\mathrm{bd}$ is the BD radius). We will hereafter assume that all BDs have $R_\mathrm{bd}=0.1 \, R_\odot$. For a star and BD pair the capture cross section is:
\begin{equation}
\label{eq:sigma_capt}
    \sigma_\mathrm{capt}=\begin{cases}
    \sigma_\mathrm{capt}' - \sigma_\mathrm{coll}\, & \sigma_\mathrm{capt}' >\sigma_\mathrm{coll}\\
    0   \, &\rm{otherwise}
    \end{cases},
\end{equation} where $\sigma_\mathrm{coll}$ is the collisional cross section (including gravitational focusing) and 
\begin{equation}
\label{eq:sigmacapt_dash}
\sigma_\mathrm{capt}' = \pi a_\mathrm{capt}^2 \left[ 1 + \frac{2Gm_*(1+q)}{v_\infty^2 a_\mathrm{capt}}\right]
\end{equation} is the capture cross section if collisions are ignored. 

Within the \textsc{Mocca} framework, we implement the tidal capture scenario in the same way as stellar collisions \citep{Freitag02}. In brief, this involves looping over all stars within a local subset \citep[`zone' -- see][]{Gie98}, and finding a corresponding BD pair at random. {We compute the local number density $n_\mathrm{bd}$ of BDs in the same way as stars in the \textsc{Mocca} framework. In brief, this involves finding a number of the closest objects in cluster radius ($r$) space and normalising by the minimum spherical shell volume that encloses them.} The probability of capture between the pair is:
\begin{equation}
    P_{\mathrm{capt}} = n_\mathrm{bd} \sigma_\mathrm{capt} v_\infty \Delta t, 
\end{equation} for time-step $\Delta t$. In this way, due to the normalisation by the BD density, it is only necessary to loop over all stars and not both stars and BDs. In the case of capture, a star-BD binary is produced with a circular orbit (assuming a  short circularisation time-scale) and semi-major axis $a_\mathrm{bd} = 2 a_\mathrm{p}$ \citep{Mardling96}. {To determine $a_\mathrm{p}$ for a given encounter, we first draw the encounter cross section uniformly between $\sigma_\mathrm{coll}$ and $\sigma'_\mathrm{capt}$, then assign the corresponding $a_\mathrm{p}$. Apart from for tidal capture scenarios, BDs are treated as main sequence stars in \textsc{Mocca}, with associated collision probabilities and dynamical interactions.}

\subsection{Capture rates}
\label{sec:theory_rates}

{As discussed in Section~\ref{sec:post-process}, it is useful to not only compute the cross section and capture probability for BD-star pairs in the Monte Carlo simulation but also post-process the encounter probability for individual stars. This allows us to compute the scaling of the capture probability over the physical parameter space, where the Monte Carlo may only yield a small number of encounters that result in large uncertainties. Therefore we compute the encounter rates as a function of stellar and environmental parameters in this section. We make an analytic estimate assuming no collisions to give an intuition as to how the capture rate scales (Section~\ref{sec:analytic_approx}). We then compute the capture rates with star-BD collisions included, demonstrating how these collisions reduce the capture rates (Section~\ref{sec:full_calc}).  } 

\begin{figure}
    \centering
   \subfloat[\label{subfig:sigvar}Encounter cross sections with relative speed]{\includegraphics[width= 0.8\columnwidth]{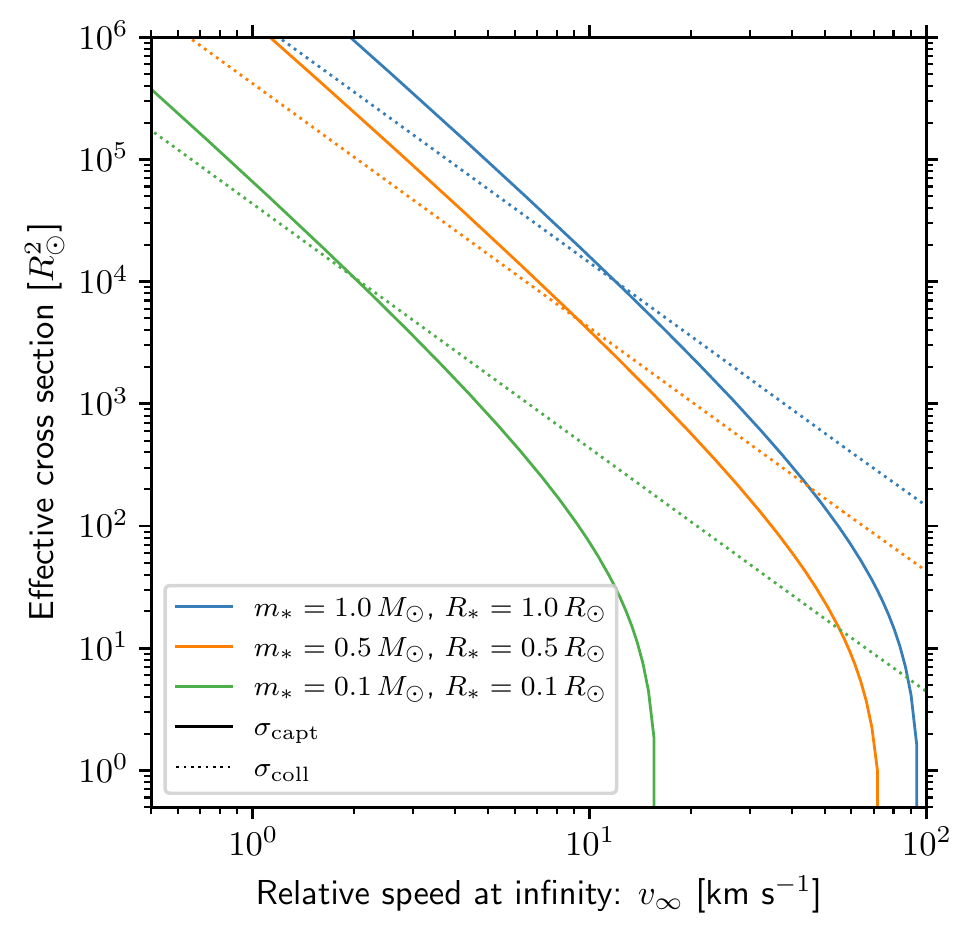}}\\\vspace{-10pt}
   \subfloat[\label{subfig:dgcdv}Differential capture rate]{\includegraphics[width= 0.8\columnwidth]{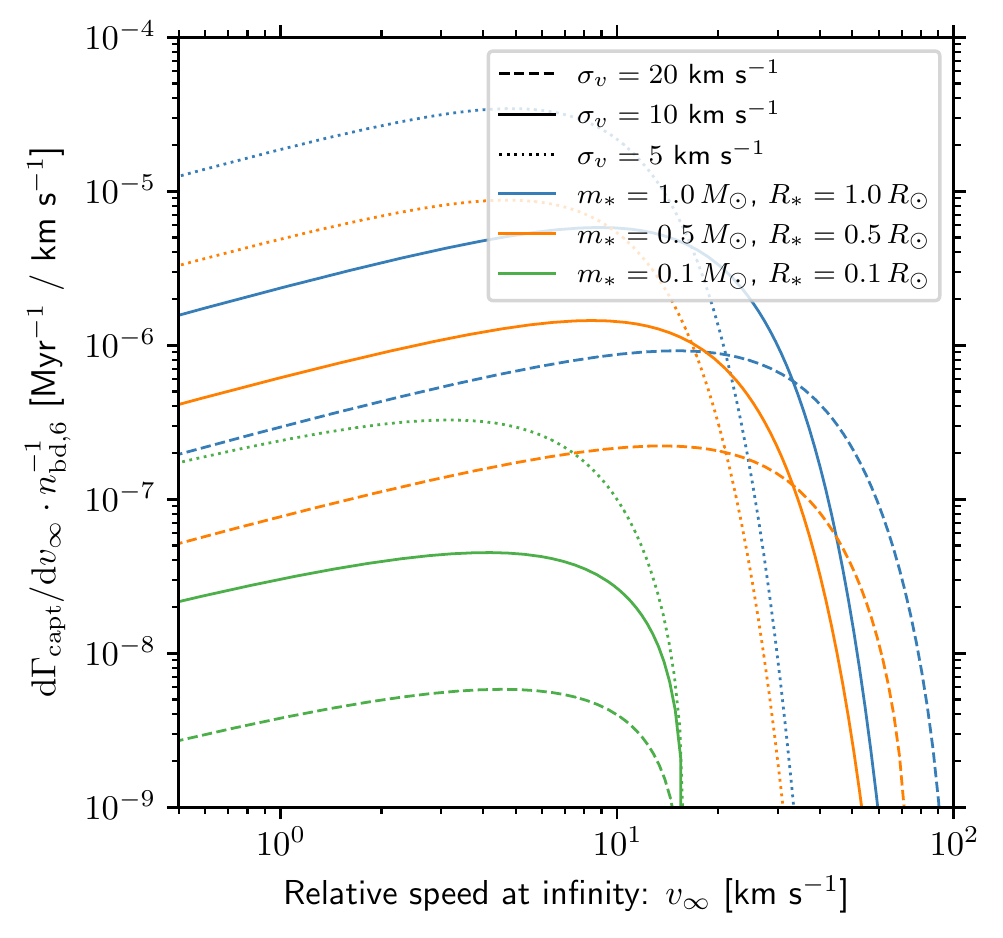}}\\ \vspace{-10pt}
        \subfloat[\label{subfig:Gamma_capt}Overall per star capture rates ]{\includegraphics[width= 0.8\columnwidth]{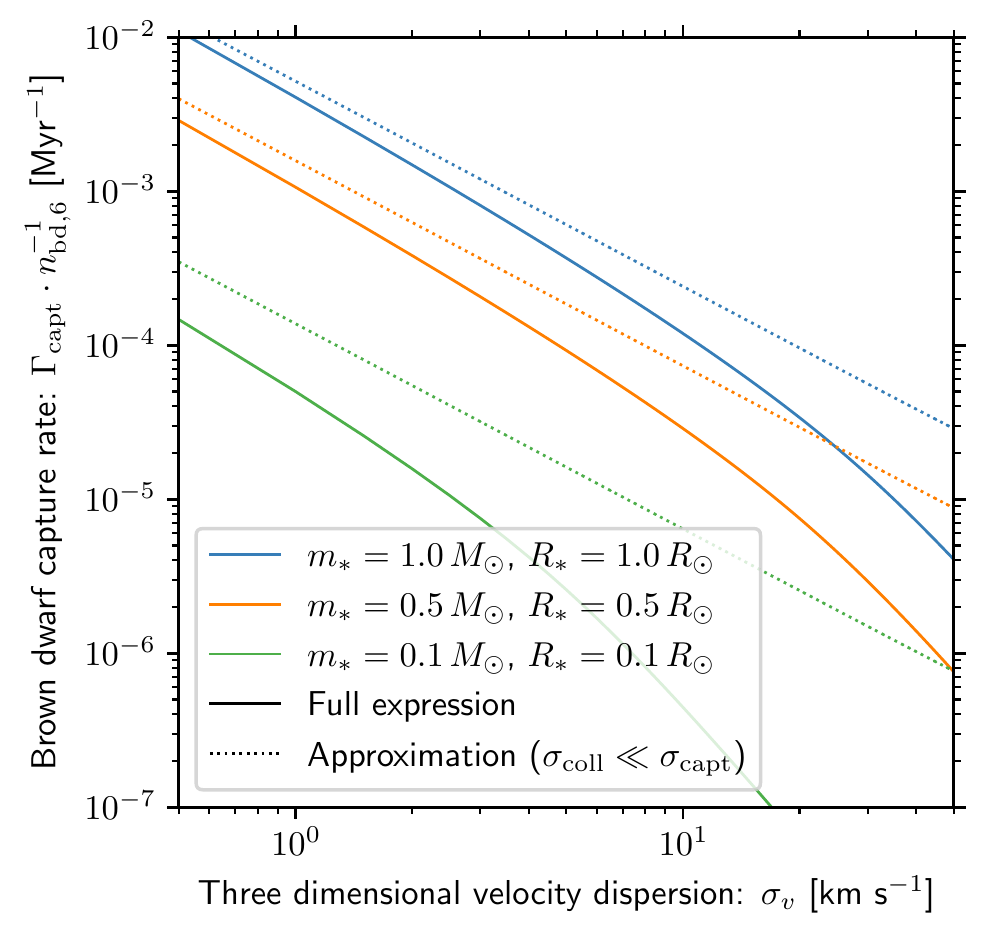}}\\\vspace{-3pt}
    \caption{Computations of the key quantities in determining the tidal capture rates of BDs in a given stellar environment. Figure~\ref{subfig:sigvar} shows the effective cross sections of collision (dotted lines) and capture (solid lines) for varying relative speeds at infinite separation. Figure~\ref{subfig:dgcdv} shows the corresponding differential capture rates (equation~\ref{eq:Gamma_capt}) for different velocity dispersions. Figure~\ref{subfig:Gamma_capt} shows the integrals of this differential across all velocities (solid lines) compared to the approximation in equation~\ref{eq:approx_captrate} (dotted lines). In all cases, the lines are coloured by the assumed stellar properties. Results in Figures~\ref{subfig:dgcdv} and~\ref{subfig:Gamma_capt} are shown for BD number density $n_\mathrm{bd}/10^6$~pc$^{-3} = n_{\mathrm{bd},6} = 1$ and scale linearly with this value. }
    \label{fig:Gammacapt_sigv}
\end{figure}

\subsubsection{Estimated encounter rate by \citetalias{Bon03}}

 {Given a local velocity dispersion and BD density, the instantaneous tidal capture rate for an individual star can be estimated. This rate $\Gamma_\mathrm{capt}$ is the rate at which neighbours pass within the effective cross section for capture, equation~\ref{eq:sigma_capt}. In general, if all objects have mass $m$, velocity dispersion $\sigma_v$ and density $n$ then the encounter rate within a given radius $a_\mathrm{enc}$ can be written \citep{Binney08}:}
 \begin{equation}
     \Gamma_\mathrm{enc} = 16\sqrt{\pi} \cdot n \sigma_v a_\mathrm{enc}^2 \cdot \left( 1+\frac{G m}{2\sigma_v^2 a_\mathrm{enc}}\right).
 \end{equation}{In the gravitationally focused regime, the second term in brackets dominates and $\Gamma_\mathrm{enc} \propto a_\mathrm{enc}/\sigma_v$. If $a_\mathrm{enc} = a_\mathrm{capt}- (R_*+R_{\mathrm{bd}})$ is not strongly dependent on $v_\infty$, then for BD density $n_\mathrm{bd}$ the BD capture rate for a star of mass $m_*$ is:
\begin{multline}
\label{eq:Gammacapt_Bonnell}
    \Gamma_\mathrm{capt}^\mathrm{BCB+03} \approx 1.4 \times 10^{-4} \, \left( \frac{n_\mathrm{bd}}{10^6\, \rm{pc}^{-3}}\right) \times \\ 
    \times \left( \frac{\sigma_v}{10\, \rm{km\, s}^{-1}}\right)^{-1} \frac{a_\mathrm{enc}}{1\, R_\odot} \frac{m_*}{1\, M_\odot} \,\rm{Myr},
\end{multline}where \citetalias[][]{Bon03} apply this expression with fixed $v_\infty = \sigma_v = 10$~km/s. Adopting equation~\ref{eq:Gammacapt_Bonnell} and assuming $v_\infty = \sigma_v$ to obtain $a_\mathrm{enc}$ is accurate when the $\sigma_v$ is sufficiently small  -- i.e. when $\sigma_\mathrm{capt}' \gg \sigma_\mathrm{coll}$ for $v_\infty \approx \sigma_v$. However, both $\sigma'_\mathrm{capt}$ and $\sigma_\mathrm{coll}$ are dependent on the relative velocities of the pairs, stellar mass and radius in different ways, such that understanding the scaling of $\sigma_\mathrm{capt}$ on these properties is non-trivial when $\sigma_\mathrm{capt}' \approx \sigma_\mathrm{coll}$ for $v_\infty \approx \sigma_v$. It is therefore unclear what kind of encounters dominate the overall capture rate in general and we will find that $\sigma_v$ significantly exceeds $10$~km/s in our dynamical model for 47 Tuc. For this reason, we must generally integrate over the full differential capture rate, including the capture cross section. }

\subsubsection{Analytic approximation without collisions}
\label{sec:analytic_approx}
The differential capture rate as a function of the velocity at infinity $v_\infty$ is:
\begin{equation}
\label{eq:Gamma_capt}
   \mathrm{d} \Gamma_\mathrm{capt} = \sigma_\mathrm{capt}(v_\infty; q, R_*) \,n_{\rm{bd}} \,  v_\infty g(v_\infty; \sigma_v) \,\rm{d}v_\infty  
\end{equation}where
\begin{equation}
    g(v_\infty; \sigma_v) = \frac{v_\infty^2}{2\sqrt{\pi} \sigma_v^3} \exp \left(\frac{-v_\infty^2}{4\sigma_v^2} \right)
\end{equation} is the Maxwell-Boltzmann distribution, or the relative asymptotic speed $v_\infty$ distribution for dispersion $\sigma_v$. In the limit $\sigma_\mathrm{capt}\gg \sigma_\mathrm{coll}$ (small $v_\infty$, large $m_*$) for the dominant capture scenarios, equation~\ref{eq:Gamma_capt} can be integrated over all velocities to yield the analytic upper-limit to the capture rate:
\begin{multline}
\label{eq:approx_captrate}
    \Gamma_\mathrm{capt} < 3.7 \times 10^{-7}  \left[\left(\frac{R_*}{1\,R_\odot}\right)^5 \frac{m_*}{1\,M_\odot}  \frac{\sigma_v}{10\,\rm{km \, s}^{-1} }q (1+q) \right]^{1/3}\times \\
    \times \left(1+ \phi_\mathrm{grav} \right) \cdot \left( \frac{n_{\rm{bd}}}{10^6\, \rm{pc}^{-3}}\right) \, \rm{Myr}^{-1},
\end{multline} where 
\begin{align}
\nonumber
    \phi_{\rm{grav}} &\approx 423\, q^{-1/6}(1+q)^{5/6} \quad\times \\ &\qquad\times \left(\frac{m_*}{1\,M_\odot}\right)^{5/6} \left(\frac{R_*}{1\,R_\odot}\right)^{-5/6} \left(\frac{\sigma_v}{10\, \rm{km\,s}^{-1}} \right)^{-5/3}
\end{align} is the gravitational focusing factor. While these expressions are approximate, they highlight three things:
\begin{enumerate}
    \item The relevant encounters are in practice always gravitationally focused ($\phi_\mathrm{grav}\gg1$), independently of the type of star under consideration, where we assume $R_*\propto m_*^\beta$ with $\beta\sim 1$. The velocity dispersion required for $\phi_\mathrm{grav}\sim 1$ is $\sigma_v \sim 100$~km~s$^{-1}$. 
    \item The capture rate scales steeply with the mass (radius) of the star: $\Gamma_\mathrm{capt} \propto q^{1/6} R_*^{5/6} m_*^{7/6} \propto m_*^{1 +5\beta/6}$. This steep scaling highlights the importance of the mass function in assessing the total number of encounters. In particular, for mass function $\xi$ we have $\Gamma_\mathrm{capt}\cdot \xi \,\mathrm{d}m_* \propto m_*^{-\gamma}\,\mathrm{d}m_*$ for $\gamma \lesssim 1$, such that the integral diverges for large $m_*$. The overall capture rate is therefore initially dependent on the choice of maximum stellar mass $m_\mathrm{max}$.
    \item The capture time-scale ($\propto 1/\Gamma_\mathrm{capt}$) scales super-linearly with the velocity dispersion ($\Gamma_\mathrm{capt} \propto \sigma_v^{-4/3}$). The temporal and spatial evolution of the local velocity dispersion within a globular cluster is therefore an important factor in determining capture frequency.
\end{enumerate}

\subsubsection{Full calculation with collisions}
\label{sec:full_calc}

The stellar mass and velocity dispersion become even  more important  when one numerically integrates the full expression for equation~\ref{eq:Gamma_capt}. {We show the relevant quantities in computing the capture rate in Figure~\ref{fig:Gammacapt_sigv}. The comparison of the cross sections for capture and collisions are in Figure~\ref{subfig:sigvar}. The collision cross section exceeds the cross section for capture above some relative speed $v_\infty$, which increases with stellar mass. The effect of this on the differential encounter rate for given velocity dispersion $\sigma_v$ is shown in Figure~\ref{subfig:dgcdv}, the form of which is non-trivially dependent on the stellar properties. }

The analytic approximation in the limit $\sigma_\mathrm{coll} \ll \sigma_\mathrm{capt}$ (equation~\ref{eq:approx_captrate}) is compared with this numerical integration in Figure~\ref{subfig:Gamma_capt} as a function of the velocity dispersion. As $\sigma_v$ becomes large, from equation~\ref{eq:capt_rad} we have that $a_\mathrm{capt}/R_*$ becomes small for all possible encounters, such that neglecting collisions is not possible. The result is a (stellar mass dependent) steep decline in the capture rate with $\sigma_v$, much steeper than the analytic approximation of $\Gamma_\mathrm{capt}\propto \sigma_v^{-4/3}$. Since the approximation is only valid for smaller $\sigma_v$ than is typical for globular clusters, we will hereafter always adopt the full numerical integration when estimating capture rates. 

The deviation from our approximation particularly affects the lower mass stars, for which gravitational focusing becomes weaker at a lower velocity dispersion relative to higher mass stars. The decrease in capture efficiency for low mass stars is precipitous for $m_*\lesssim 0.5\,M_\odot$, such that the stellar mass function must be considered in order to compute global capture rates. We further consider how the evolving mass function alters the overall capture rates in Appendix~\ref{app:massfunc}, where we justify adopting $m_* = 0.7\,M_\odot$ to approximate the global capture rate over the lifetime of 47 Tuc.

\section{Dynamical modelling}

\label{sec:Num_Method}

\subsection{Monte Carlo simulations}

Simulating the dynamical evolution of globular clusters directly using $N$-body calculations over their $\sim 10$~Gyr evolution is not computationally practicable. For this reason, the \textsc{Mocca} code \citep{Gie98, Gie01} has been developed as a {Monte Carlo} approach to statistically computing the evolution of massive, dense stellar clusters by solving the Fokker--Planck equation \citep[see also][]{Sto82}. This approach allows a fast and accurate calculation of the dynamical evolution of the stellar population in a globular cluster over its lifetime. The added bonus of using this code is that it has already been applied to model the evolution of 47 Tuc \citep[][hereafter \citetalias{Gie11}]{Gie11}. We are therefore able to adopt the parameters obtained in this previous modelling effort. Where appropriate, we make similar comparisons to observational constraints as \citetalias{Gie11}. \textsc{Mocca} has since been updated to incorporate the \textsc{Fewbody} code \citep{Fregeau04, Fregeau12} into an improved prescription for interactions between multiple systems, as described by \citet{Hyp13}. Stellar evolution modules by \citet{Hurley00, Hurley02} are used to compute single star and binary evolution. We do not include tidal forces between a star and a companion in this evolution.  

\subsection{Initial conditions for 47 Tuc}

\label{sec:ICs}
\begin{figure*}
 \subfloat[\label{subfig:surfmag}Surface brightness]{\includegraphics[width= 0.47\textwidth]{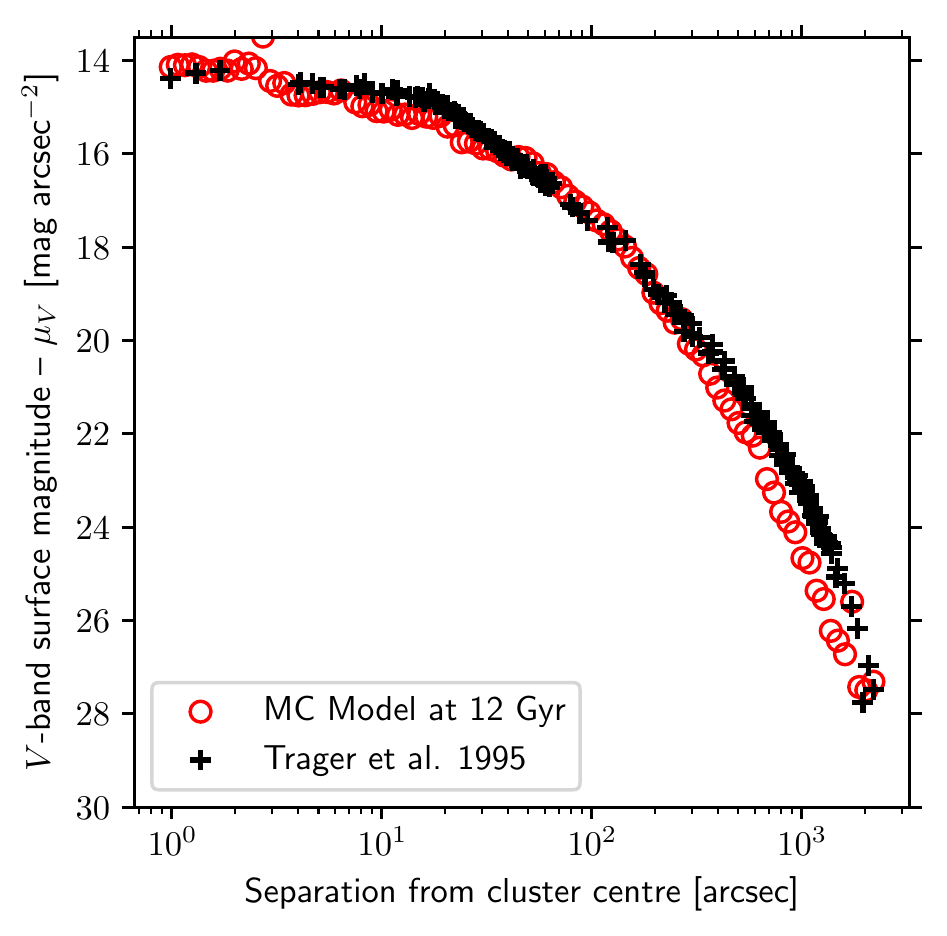}}
 \subfloat[\label{subfig:los_vdisp}Line of sight velocity dispersion]{\includegraphics[width= 0.47\textwidth]{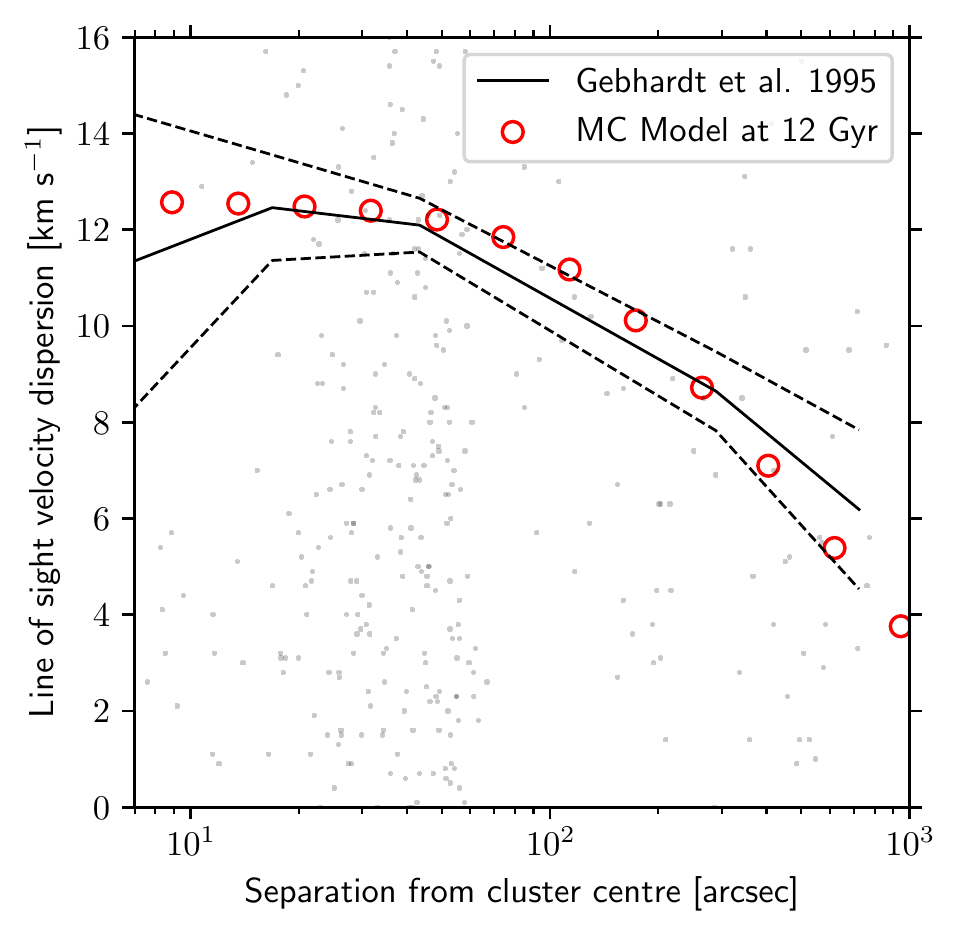}}
       
     \caption{Observational constraints on the dynamical properties of 47 Tuc compared to the results of our \textsc{Mocca} (Monte Carlo) model with the parameters listed in Table~\ref{table:dyn_mods}. Figure~\ref{subfig:surfmag} shows the visible surface brightness profile of the Monte Carlo model (red circles) compared to the observed profile found by \citet[][black crosses]{Tra95}. The line-of-sight velocity dispersion is compared in Figure~\ref{subfig:los_vdisp}, where the Monte Carlo results are again red circles. The relative line of sight velocities measured by \citet{Gebhardt95} are shown by faint points and the inferred dispersion shown as a black line. The sampling uncertainties are indicated by dashed lines.  }
     \label{fig:47Tuc_obs}
\end{figure*}

\begin{table}
\centering 
 \begin{tabular}{c c c } 
 \hline
 Parameter & Property & Value     \\ [0.5ex] 
 \hline
$N_*$ & Number of stars &  $2 \times 10^6$ \\
$N_\mathrm{bd}$& Number of  BDs  &  $2\times 10^6$ \\
$N_\mathrm{bin}$ & Number of binaries &  $4.4\times10^4$ \\ 
$N_\mathrm{pl}$ & Number of planets &  $2 \times10^4$ \\ 
$W_0$ & Central concentration & $7.5$ \\
$m_\mathrm{br}/M_\odot$ & Break mass & 0.8 \\
$m_\mathrm{max}/M_\odot$ &  Max. mass  & 50\\
$\alpha_1$ &  IMF slope $m<m_\mathrm{br}$ & 0.4 \\
$\alpha_2$ &  IMF slope $m>m_\mathrm{br}$ & 2.8\\
$T_\mathrm{age}/$Gyr & Age & $12$ \\
$Z/Z_\odot$ & Metallicity (dex) & -0.6 \\
$e_0$ & Planet eccentricity &  0.9 \\ 
$a_0/$au & Planet semi-major axis &  $5$\\ 
 \hline
\end{tabular}
\caption{Initial condition parameters used for the Monte Carlo globular cluster model discussed in the text. Where appropriate, choices are made to match the model of \citetalias{Gie11} for 47 Tuc. BDs and planets are the same except that planets are initially companions to stars.}
\label{table:dyn_mods}
\end{table}

The initial conditions that we adopt are motivated by the findings of \citetalias{Gie11}, who reproduced the key observable properties of 47 Tuc. The main parameters are summarised in Table~\ref{table:dyn_mods}. The binary population are drawn from a log-uniform distribution between $1{-}100$~au. {The initial conditions include a number of choices that are somewhat artificial (such as the low maximum stellar mass, small binary fraction and no mass fallback for black hole formation). These choices were invoked to reproduce the unusual surface brightness and velocity dispersion profiles. A low binary fraction is also convenient in our context, because we compute tidal capture rates only between BDs and single stars. More recent models incorporating a higher binary fraction and maximum stellar mass can reproduce the frequency of special objects (e.g. black holes binaries and pulsars) and central surface brightness, but do not presently reproduce the observed surface brightness profile (A. Askar -- private communication). The most important property for computing the BD capture rate is the stellar density profile. We therefore retain the parameters of \citetalias{Gie11} with which we find good agreement with the observed density and velocity dispersion profiles (see Section~\ref{sec:47Tuc_obs_comp}). Provisional checks using the unpublished alternative models yield BD capture rates similar to those we obtain with our fiducial model. However, we emphasise initial conditions that can reproduce the properties of globular clusters are degenerate. Although models that yield a similar density profile probably yield similar capture rates (see discussion of caveats in Section~\ref{sec:caveats}), an extensive parameter study investigating these choices is outside of the scope of this work.}


We additionally include a BD population, with equal numbers as the stars. The masses of the BDs have initial masses $m_\mathrm{bd} = 0.079\, M_\odot$, to ensure that their masses are lower than that of the least massive stars. In the case that an object has a mass that exceeds $0.08 \, M_\odot$ (for example, via a collision/merger), then the object is no longer defined as a BD. The initial spatial distribution of BDs is assumed to be the same as the stellar population (i.e. no primordial mass segregation).

Finally, we add a population of `migrating planets' around 1 percent of the initial stellar population. {The orbital evolution of this population will be considered in Paper II, but are not relevant in this work.} We adopt the same mass as the BDs for simplicity. While this results in a greater mass ratio $q$ than for planets, the scattering cross section for binaries is only weakly dependent on the mass ratio, particularly for $q\lesssim 0.1$ \citep{Fregeau04}. We initialise all of the planet orbits to have eccentricity $e=0.9$ and semi-major axes $a = 5$~au, reflecting a Jupiter analogue with high eccentricity. Planets are paired with stars drawn from the same IMF as single stars. {Because the companion (planet) population is a small fraction of the stellar population, as well as low mass and with small semi-major axis compared to the binaries, this population has a negligible influence on the overall evolution of the \textsc{Mocca} simulation (confirmed by performing runs without them).  We do not include these systems in the tidal capture Monte Carlo routine, although this too has a negligible effect on the overall capture rate due to the low companion fraction. They are treated as binaries within the \textsc{Mocca} framework, undergoing single-binary and binary-binary interactions integrated with \textsc{Fewbody}. We will only refer to this population again in Paper II. }

\subsection{Comparison with observed properties of 47 Tuc}
\label{sec:47Tuc_obs_comp}
We wish to ensure that the model approximately reproduces the key physical properties of 47 Tuc at its present age. These are the stellar density and velocity dispersion profile. \citetalias{Gie11} fitted their models to the surface brightness profile as measured by \citet{Tra95} and the line-of-sight velocity dispersion inferred from the measurements of \citet{Gebhardt95}. For direct comparison, we perform the same comparisons for our model. 

The $V$-band surface magnitude profile after integrating the model for 12 Gyr is shown in Figure~\ref{subfig:surfmag}, adopting a distance of $4$~kpc. The profile is calculated by averaging over the $V$-band luminosity contribution of concentric shells of stars for distance $d$ from the cluster centre:
\begin{equation}
\label{eq:Vband_surf}
   \Sigma_V(d) =\sum_{r_i>d}\frac{L_V}{2\pi r_i^2} \frac{r_i}{\sqrt{r_i^2-d^2}}.
   \end{equation}
   We then convert this to a surface magnitude by the expression:
\begin{equation}
    \mu_V = V_\odot -2.5\log \Sigma_{V}' + A_V,
\end{equation}where $V_\odot = 4.80$ is the solar $V$-band magnitude and $\Sigma_{V}' $ is $\Sigma_{V} $ in units of solar $V$-band luminosity per square arcsecond. {We assume small visual extinction $A_V$. Commonly, the \citet{Harris96} value of $E(B-V) = 0.04$ is used \citep[although smaller value $E(B-V)=0.024$ may also be adopted --][]{Crawford75}, to give $A_V \approx 0.12$ for relative visibility $R_V = 3.1$.} Despite the updated version of \textsc{Mocca} and the inclusion of BDs compared to \citetalias{Gie11}, we find reasonable agreement between the model and the observed profile obtained by \citet{Tra95} similarly to \citetalias{Gie11}. 

The line-of-sight velocity dispersion profile can be computed in an analogous fashion to the surface brightness and is shown in Figure~\ref{subfig:los_vdisp}. The dispersion at projected separation $d$ from the centre is the contribution of the projected contributions of the radial and tangential velocities ($v_r$ and $v_{\rm{t}}$ respectively):
\begin{equation}
    \sigma_{v,\rm{los}}^2(d)=   \frac{1}{n_{d}}\sum_{r_i>d} \frac{d}{r_i\sqrt{r_i^2-d^2}}\left[v_r^2\frac{r_i^2-d^2}{r_i^2} + \frac {v_{\rm{t}}^2} 2 \frac{d^2}{r_i^2} \right] , 
\end{equation} where 
\begin{equation}
    n_d=  \sum_{r_i>d} \frac{d}{r_i\sqrt{r_i^2-d^2}}.
\end{equation}The corresponding observed dispersion can be extracted from the measured line-of-sight velocities by computing the dispersion relative to the mean velocity, binned by separation from the cluster centre. The comparison in Figure~\ref{subfig:los_vdisp} shows that the two dispersions are similar across the separations with observational constraints. Given that both the surface density and velocity dispersion profiles are comparable to the observational constraints, we adopt this model without further (computationally expensive) parameter study.

\subsection{Stellar density and velocity evolution}

The main quantities of interest for computing the rate of close encounters in a stellar cluster are the local number density and velocity dispersion. We post-process the output of our Monte Carlo integration to track the density and velocity dispersion evolution over the 12 Gyr lifetime of 47 Tuc. We divide the stars by cluster radii into $30$ log-uniformly spaced bins between $10^{-1.5}$ and $10^1$~pc, then normalise the number in each shell by the volume to yield the density. For velocity, we perform a similar binning but then compute the dispersion $\sigma_v$ in the one dimensional velocities: $\sqrt{ v_r^2+v_\mathrm{t}^2}$, where $v_r$ is the velocity in the radial direction, and $v_\mathrm{t}$ is the tangential component (combined azimuthal and polar). 

\subsubsection{Density evolution}

\begin{figure} 
    \centering
    
    \includegraphics[width=\columnwidth]{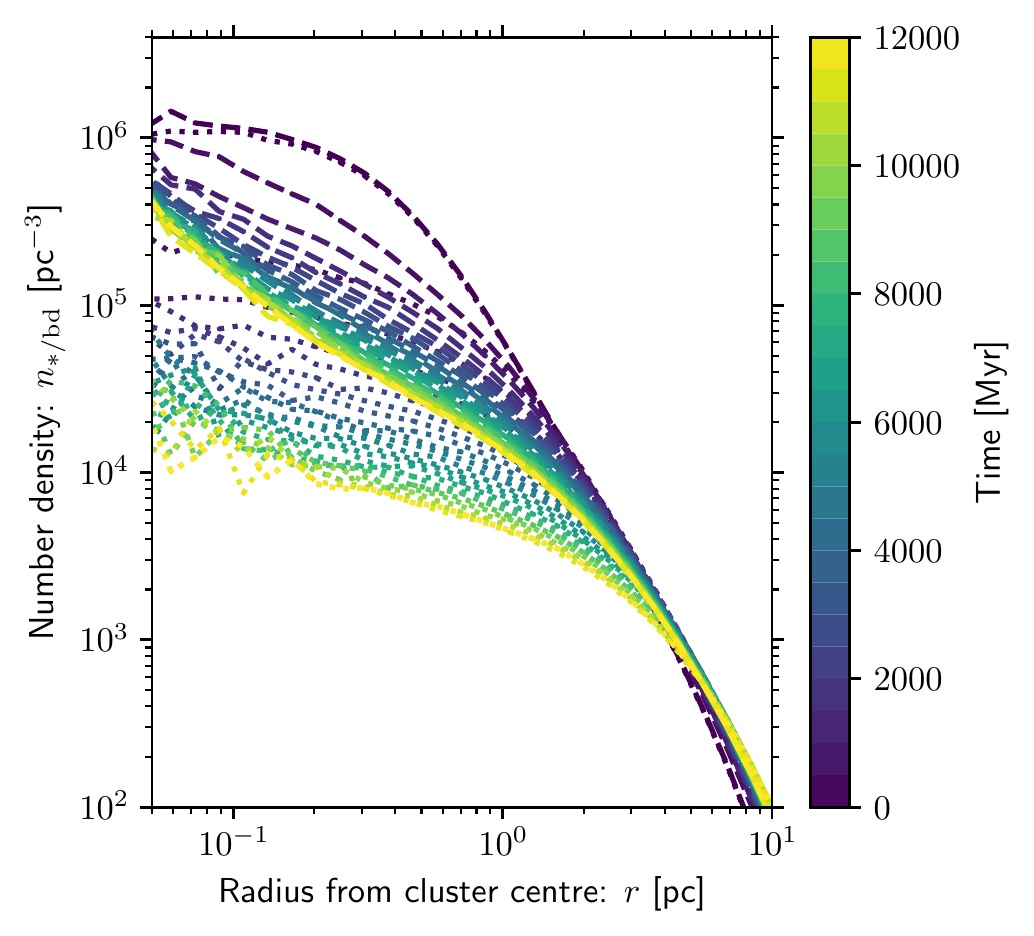}
    \caption{Density profile evolution for stars (dashed) and BDs (dotted) in our \textsc{Mocca} model for the dynamical evolution of 47 Tuc. The lines are shown every $500$~Myr, coloured by the time in the simulation. Both stars and BDs initially have the same density profile.}
    \label{fig:n_evol}
\end{figure}

We show the density profile evolution in Figure~\ref{fig:n_evol}, separated into stars (dashed) and BDs (dotted). The initial core stellar density distribution is $n_*\sim 10^6$~pc$^{-3}$ within $\sim 0.5$~pc, and stellar densities $\gtrsim 10^5$~pc$^{-3}$ are retained for several Gyr. However, this is not true for the BD population. During relaxation, dynamical interaction leads to energy equipartition, resulting in lower masses being pushed to regions of a shallower gravitational potential; this is known as mass segregation \citep{Binney08}. By this process highest density regions at the cluster centre are quickly vacated of BDs to yield densities $n_\mathrm{bd}\lesssim 10^5$~pc$^{-3}$ within $\sim 1$~Gyr. This will have significant consequences on the efficiency of tidal capture over the lifetime of 47 Tuc  (see Section~\ref{sec:bd_capt}).

\subsubsection{Velocity dispersion evolution}
\label{sec:sigmav_evol}

\begin{figure}
    \centering
    \includegraphics[width=\columnwidth]{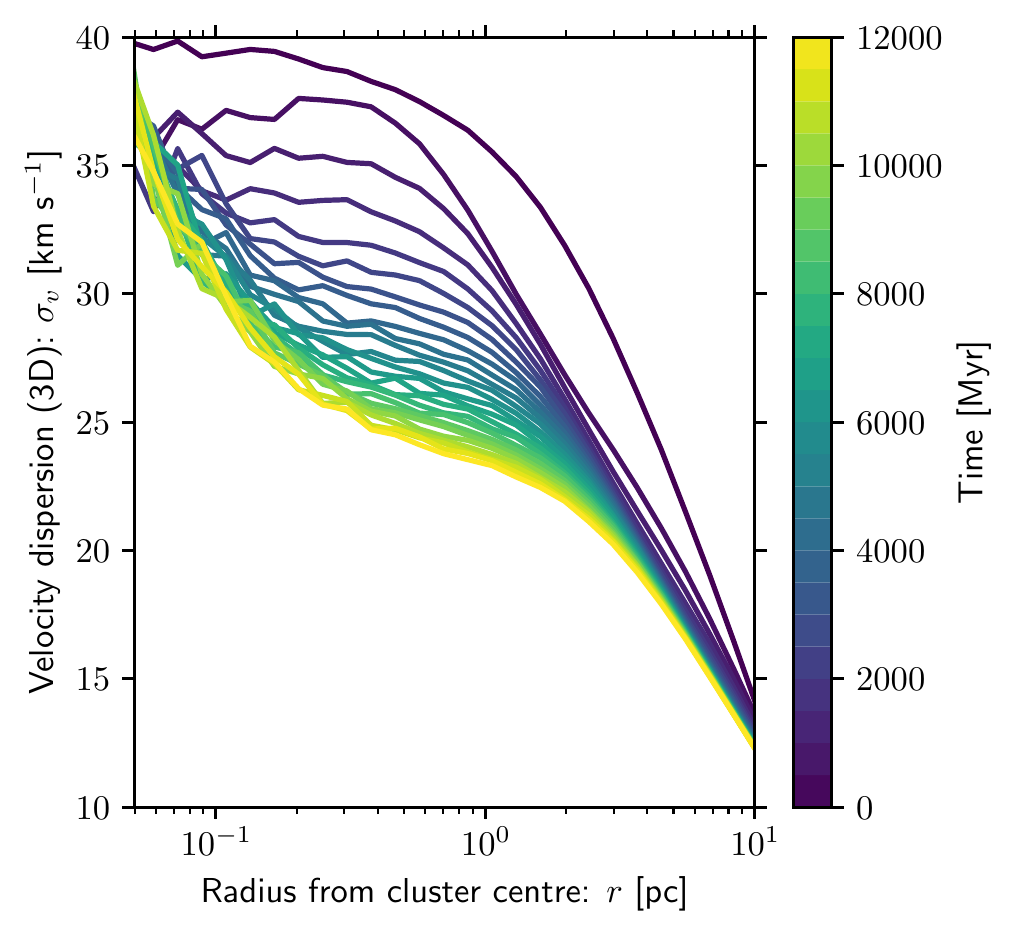}
    \caption{Velocity dispersion profile evolution in the \textsc{Mocca} model for the dynamical evolution of 47 Tuc. Lines are coloured by time in the simulation, each separated by $500$~Myr.}
    \label{fig:sigmav_evol}
\end{figure}

{The evolution of the velocity dispersion profile is shown in Figure~\ref{fig:sigmav_evol}. In this case the velocity dispersion across the majority of the cluster decreases as it relaxes. The very inner region retains a high velocity dispersion, which is due to the fact that the majority of trajectories passing through this region are at the pericentre of an eccentric orbit. The physical (three dimensional) dispersion is a factor $\sqrt{3}$ greater than the one dimensional dispersion. However, the line-of-sight dispersions inferred from radial velocity measurements and our reconstructed `observation' of the model ($\sim 10{-}12$~km~s$^{-1}$ in the central regions; Section~\ref{sec:ICs}) are lower again by a factor of order unity. This is due to projection effects. The stellar density profile means that the majority of stars are a few parsec from the centre in three dimensions. When one infers the dispersion of radial velocities at projected (two dimensional) separations smaller than this, the measurements are mostly made for outer, lower velocity stars that fall along the line-of-sight. Thus the apparent dispersion is an underestimate of the physical one dimensional dispersion in the central regions. }


As a result of these considerations, the inner physical three dimensional dispersion is much larger in the highest density regions than the fiducial $10$~km~s$^{-1}$ assumed by \citetalias{Bon03}. In Section~\ref{sec:theory_rates} we show that the capture rate decreases much steeper than linearly with $\sigma_v$. These two findings demonstrate why the approach we take is necessary in computing accurate capture rates over the lifetime of a globular cluster. A velocity dispersion of $\sigma_v= 30$~km~s$^{-1}$ can result in order of magnitude decreases in the capture rate compared to $\sigma_v=10$~km~s$^{-1}$, depending on the stellar mass. The full, local velocity dispersion evolution is therefore a necessary ingredient in the calculations we perform.

\section{Results}
\label{sec:results}

\subsection{Overall capture rate}
\label{sec:bd_capt}
\label{sec:overall_capt}
\begin{figure}
    \centering
    \includegraphics[width=\columnwidth]{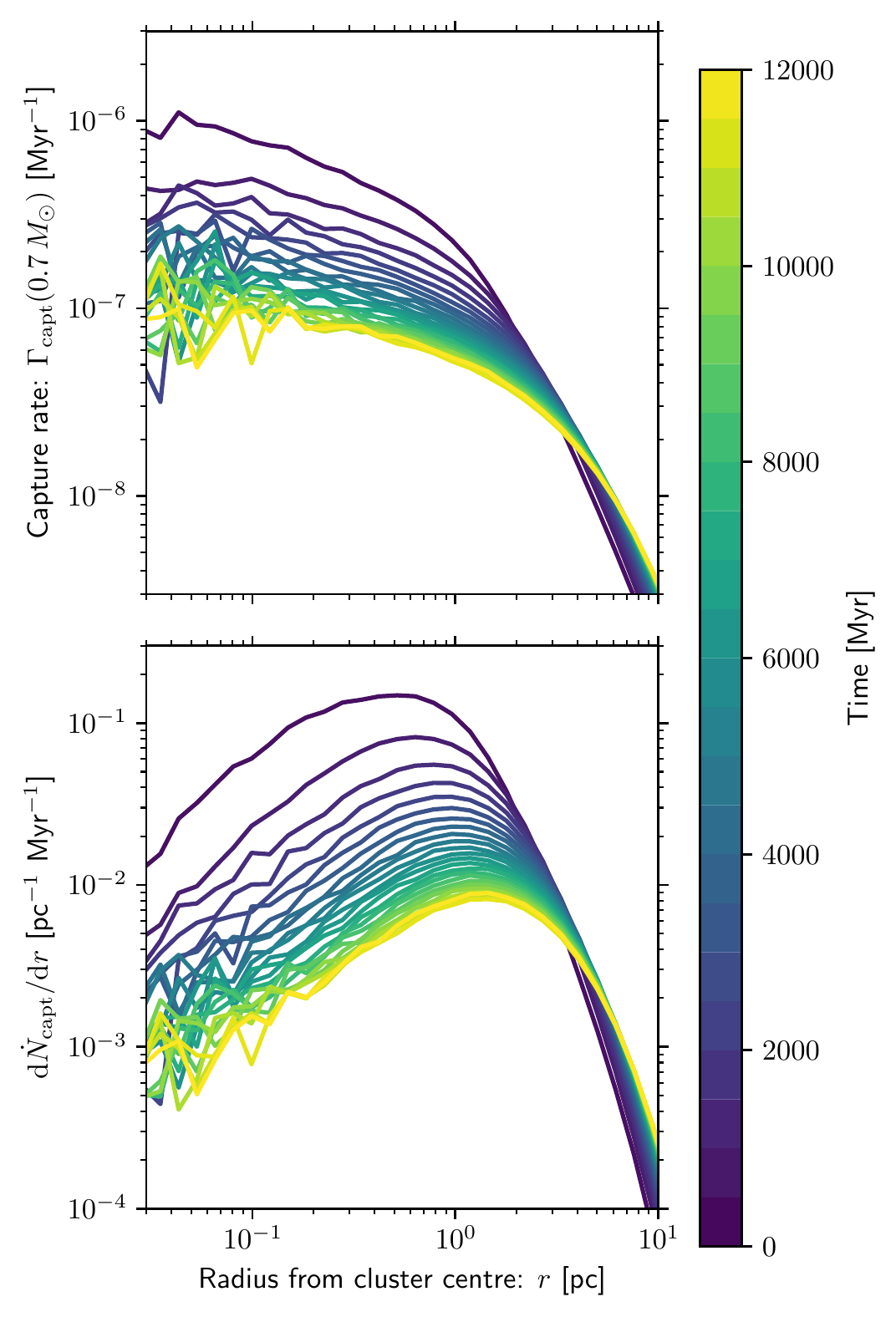}
    \caption{Top: The instantaneous rate $\Gamma_\mathrm{capt}$ of tidal BD capture for a star of mass $m_*=0.7\, M_\odot$ with the mass-radius relation given by equation~\ref{eq:approx_mr} at a given cluster radius within 47 Tuc. Contours are computed by integrating over equation~\ref{eq:Gamma_capt} given the BD number density and velocity dispersion shown in Figures~\ref{fig:n_evol} and~\ref{fig:sigmav_evol} respectively. {Bottom: The same countours but now weighted by $4\pi r^2  n_*$, giving the estimated rate of overall tidal capture per unit radius. This is an approximation for the integrand in equation~\ref{eq:totcaptrate}. }}
    \label{fig:Gammacapt_evol}
\end{figure}

\begin{figure}
    \centering
    \includegraphics[width=\columnwidth]{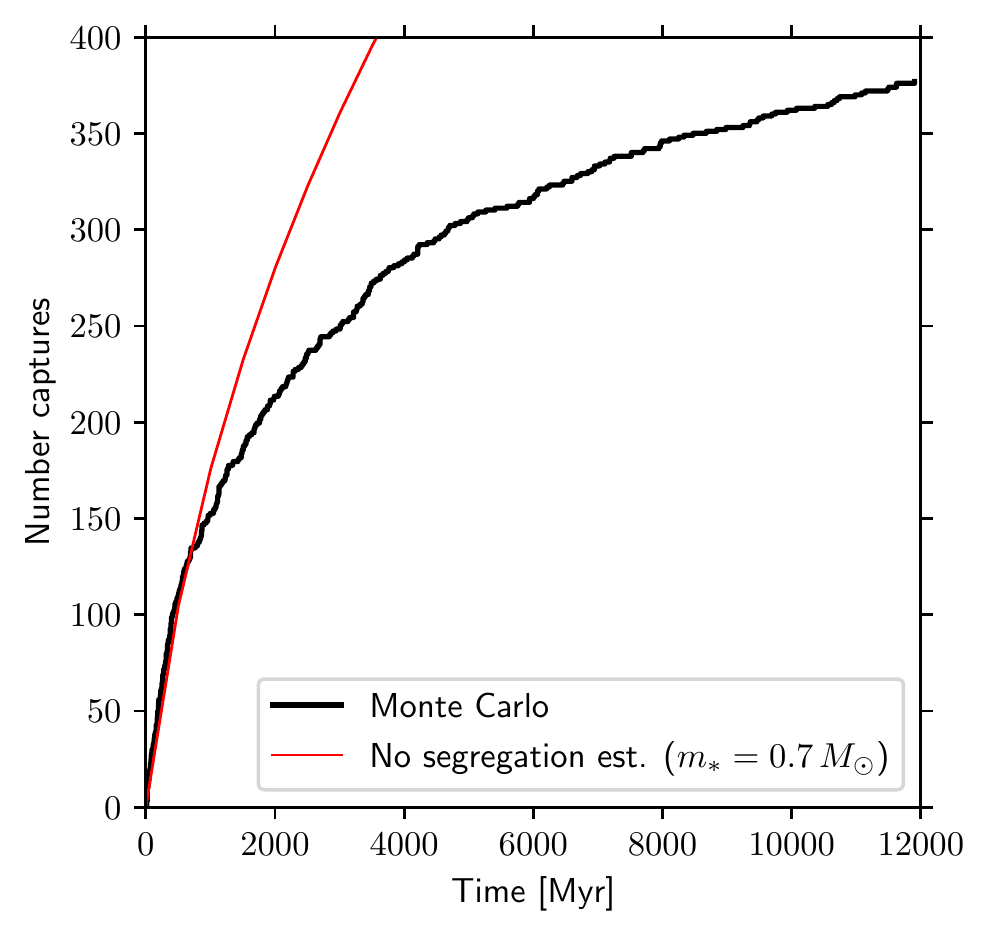}
    \caption{Cumulative number of tidal BD captures through the lifetime of 47 Tuc. The black line shows the result obtained directly from the Monte Carlo calculation. The red line is an estimate using the local stellar/BD density and velocity dispersion with equation~\ref{eq:Gamma_capt}, assuming $m_*=0.7\,M_\odot$ for all stars (see discussion in Appendix~\ref{app:massfunc}). The poor agreement after $\sim 1$~Gyr is due to {stellar} mass segregation (see text for details).}
    \label{fig:MCvest_capts}
\end{figure}

The total number of BD captures in the Monte Carlo simulation of 47 Tuc is 377. In order to interpret this directly computed capture rate, we take the density and velocity dispersion profiles from our models to approximate the global capture rates by post-processing {the density and velocity dispersion profiles to obtain the encounter rates according to equation~\ref{eq:Gamma_capt}}. {To perform this calculation, we need to adopt an expression that encapsulates the stellar mass averaged encounter rate:
\begin{equation}
     \hat{\Gamma}_\mathrm{capt} =  \int_{m_{\mathrm{min}}}^{m_\mathrm{max}}  \, \mathrm{d} m_* \, \xi(m_*)  \Gamma_\mathrm{capt} (m_*),
\end{equation}which strictly requires calculating the temporal and spatial evolution of the mass function $\xi$. However, we show in Appendix~\ref{app:massfunc} that if the mass function remains constant then after a short time-scale ($\sim 100$~Myr) the removal of the most massive stars results in $ \hat{\Gamma}_\mathrm{capt} \approx \Gamma_\mathrm{capt} (0.7\,M_\odot)$, which remains true over the majority of the lifetime of 47 Tuc. This approximates the global temporal evolution of the mass function, but not the spatial variation. We will see that the assumption of a spatially homogeneous mass function is an important omission.}

Integrating over equation~\ref{eq:Gamma_capt} yields a local capture rate over the dynamical history of 47 Tuc. The results of this calculation are shown in the top panel of Figure~\ref{fig:Gammacapt_evol}. This in turn can be integrated to give a global capture rate:
\begin{equation}
\label{eq:totcaptrate}
\dot{N}_\mathrm{capt}= \int_0^\infty\,\mathrm{d}r  \,  4\pi r^2 \,  n_*(r)  \hat{\Gamma}_\mathrm{capt} (r),
\end{equation}where we adopt the stellar mass averaged capture rate $ \hat{\Gamma}_\mathrm{capt}  \approx \Gamma_\mathrm{capt} (0.7\, M_\odot)$ (see Appendix~\ref{app:massfunc}). 

The integrand of equation~\ref{eq:totcaptrate} is shown in the bottom panel of Figure~\ref{fig:Gammacapt_evol}. The evolution of the capture rate reflects the fact that in the early stages BDs occupy the central, high density regions, but are quickly evacuated to the outer regions as the cluster becomes mass segregated. During the segregation, the capture rate profile quickly transitions from being centrally concentrated to relatively flat with radius. In general, the capture rates are slightly lower than the estimates by \citetalias{Bon03} due to the considerations discussed in Section~\ref{sec:theory_rates}; principally the strong dependence of capture efficiency on the stellar mass and local velocity dispersion. 

\begin{figure}
    \centering
    \includegraphics[width=\columnwidth]{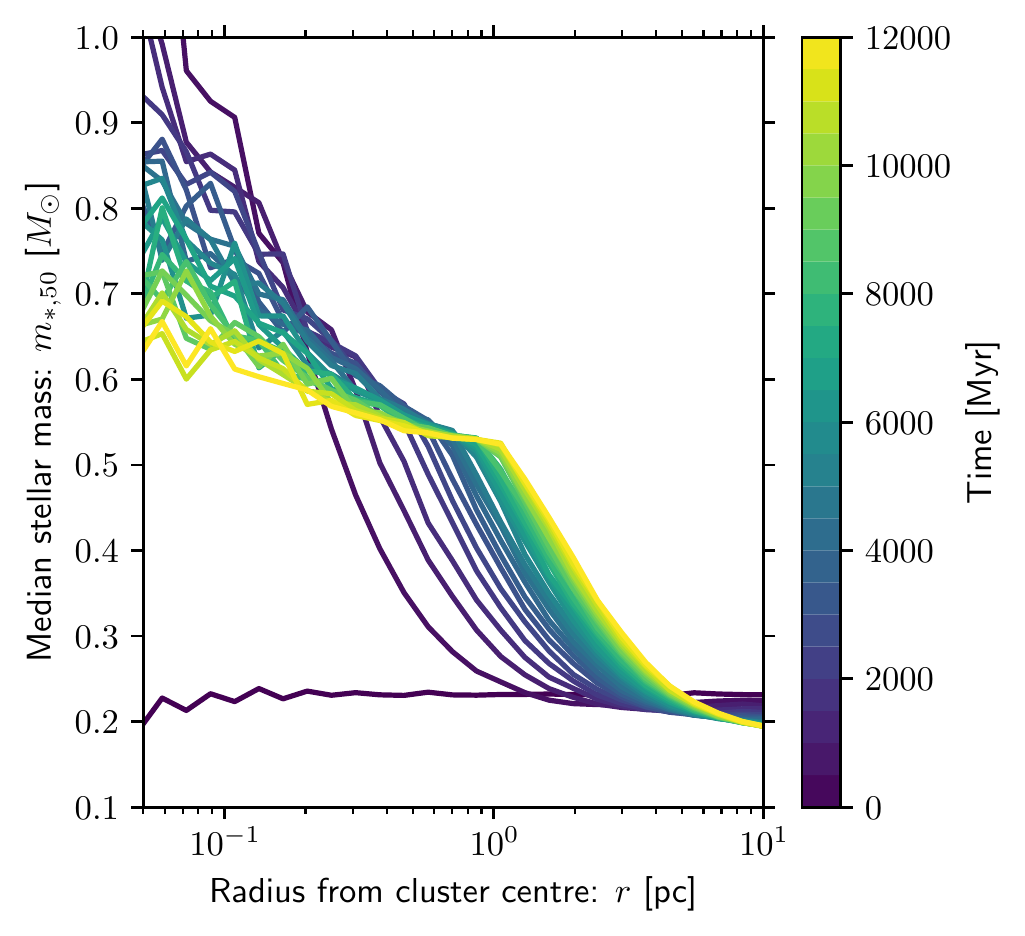}
    \caption{Median stellar mass binned by radius from the cluster centre over the lifetime of 47 Tuc in our Monte Carlo model. Lines are coloured by time in the simulation, each separated by 500~Myr.}
    \label{fig:mmed_profile}
\end{figure}

In Figure~\ref{fig:MCvest_capts} we compare the result of the estimated encounter rate computed using equation~\ref{eq:totcaptrate} {(assuming a constant stellar mass function over space and time)} with the number of captures obtained directly from the Monte Carlo simulation. The rate of capture is initially well reproduced by our estimate, but after approximately $1$~Gyr we overestimate the global rate of captures. This highlights a problem with approximating the encounter rates: we have not accounted for the segregation of \textit{stellar} mass throughout the cluster (only the BDs). Figure~\ref{fig:mmed_profile} shows the median stellar mass in radial bins over the dynamical evolution of our Monte Carlo model. The model is not initially segregated, which is why we have good agreement between the Monte Carlo and the post-processed approximation. However, a gradient in stellar masses quickly emerges as mass segregation operates. The result is that lower mass stars preferentially occupy the same regions as the BDs. Capture for low mass stars is inefficient due to the steep decline in the cross section with decreasing stellar mass when the local velocity dispersion is large (Section~\ref{sec:theory_rates}). Hence the capture rates are further suppressed by the separation of the BDs and high mass stars by which they can be efficiently captured. 

\subsection{Semi-analytic encounter rate calculations}
\label{sec:post-process}

{The Monte Carlo approach we describe in Section~\ref{sec:tid_cross} has the benefit that we can generate a realistic number of capture or scattering events across a complex parameter space. However, there are benefits to complementing the Monte Carlo with semi-analytical estimates. In particular, this allows us to more easily understand the scaling of the results with the stellar parameters and cluster properties. In the Monte Carlo calculation, the number of events may be low in some regions of parameter space (e.g. stellar mass and position), which results in large uncertainties in recovering the probability from sampling. To construct probability functions for encounter rates, a better approach is to take a subset of stars at the end of the simulation and track their encounter rate throughout their lifetime. This has the added benefit that we can use the analytic expressions to scale our results based on assumed physical properties.}

To recover the encounter rate evolution we must integrate over orbits which are much shorter than is possible to temporally resolve with the output time-step. At each snapshot we therefore recover the orbital solution by first fitting an approximate analytic double power-law density profile:
\begin{equation}
\label{eq:rho_pot}
    \rho_* = \frac{M_\mathrm{s}}{4\pi a_\mathrm{s}^3} (r/a_\mathrm{s})^{-\alpha}(1-r/a_\mathrm{s})^{\alpha-\beta}
\end{equation} to the stellar mass density, where $M_\mathrm{s}$, $a_\mathrm{s}$, $\alpha$ and $\beta$ are fitting constants. We then construct the corresponding spherically symmetric potential using the \texttt{TwoPowerSphericalPotential} class of \textsc{Galpy}\footnote{\url{http://github.com/jobovy/galpy}} \citep{Bovy15}. {Although orbits in the potential described by equation~\ref{eq:rho_pot} are not closed, since the density and velocity dispersion profiles are spherically symmetric we are only interested in the radial oscillations of the star with respect to the respect the centre of mass of the cluster. We therefore define the period $P_\mathrm{orb}$ for the star to make a single epicycle. Then the} orbitally averaged {capture} rate at the specified time-step is:
\begin{equation}
\label{eq:orbavg}
    \langle \Gamma_\mathrm{capt} \rangle (\Theta(t_\mathrm{step}))= \frac{1}{P_\mathrm{orb}} \int_{t_\mathrm{step}}^{t_\mathrm{step}+P_\mathrm{orb}} \,\mathrm{d}t \, \Gamma_\mathrm{capt}(\Theta(t)),
\end{equation}where $t$ is the time coordinate, $t_\mathrm{step}$ is the time of the snapshot and all other pertinent parameters are enclosed in $\Theta$. In practice, if the time-step between updating orbital solution $\Delta t<P_\mathrm{orb}$, then $P_\mathrm{orb}$ is replaced with $\Delta t$ in equation~\ref{eq:orbavg}. {The capture probability in equation~\ref{eq:orbavg} is computed by integrating over equation~\ref{eq:Gamma_capt} with the BD density and velocity dispersion interpolated over radius space at time $t_\mathrm{step}$.} In this way, we can estimate the probability of {capture} for a given star $i$ within a certain time $T_\mathrm{age}$:
\begin{equation}
\label{eq:Penc}
    P_{\mathrm{capt},i} =1- \exp\left[ - \int_0^{T_\mathrm{age}}\, \mathrm{d}t \,   \langle \Gamma_\mathrm{capt} \rangle ( \Theta_i(t))\right], 
\end{equation}where $\Theta_i$ are now the star-specific and time-dependent parameters that determine the {capture} rate. {In practice, we apply equation~\ref{eq:Penc} as a sum over discrete time-steps} to understand the variation in {capture} rate with the properties of the star, particularly the stellar mass.

\subsection{Brown dwarf capture and stellar mass}

{In this work, we have demonstrated the importance of stellar mass for the rates of tidal BD capture. The sample of \citetalias{Gil00} comprised $34,091$ stars in the $V$-band magnitude range $17.1 < V < 21.1$. Based on the models of \citet{Bergbusch92} for [Fe/H]$=-0.78$ at $12$~Gyr, the brightest stars included therefore have mass $m_*\approx 0.88\, M_\odot$ (approximately the turn-off from main sequence for 47 Tuc) and the lowest masses are $m_* \approx 0.52\,M_\odot$. For comparison, we must therefore consider the capture probabilities in this stellar mass range. }

\begin{figure}
    \centering
    \includegraphics[width=\columnwidth]{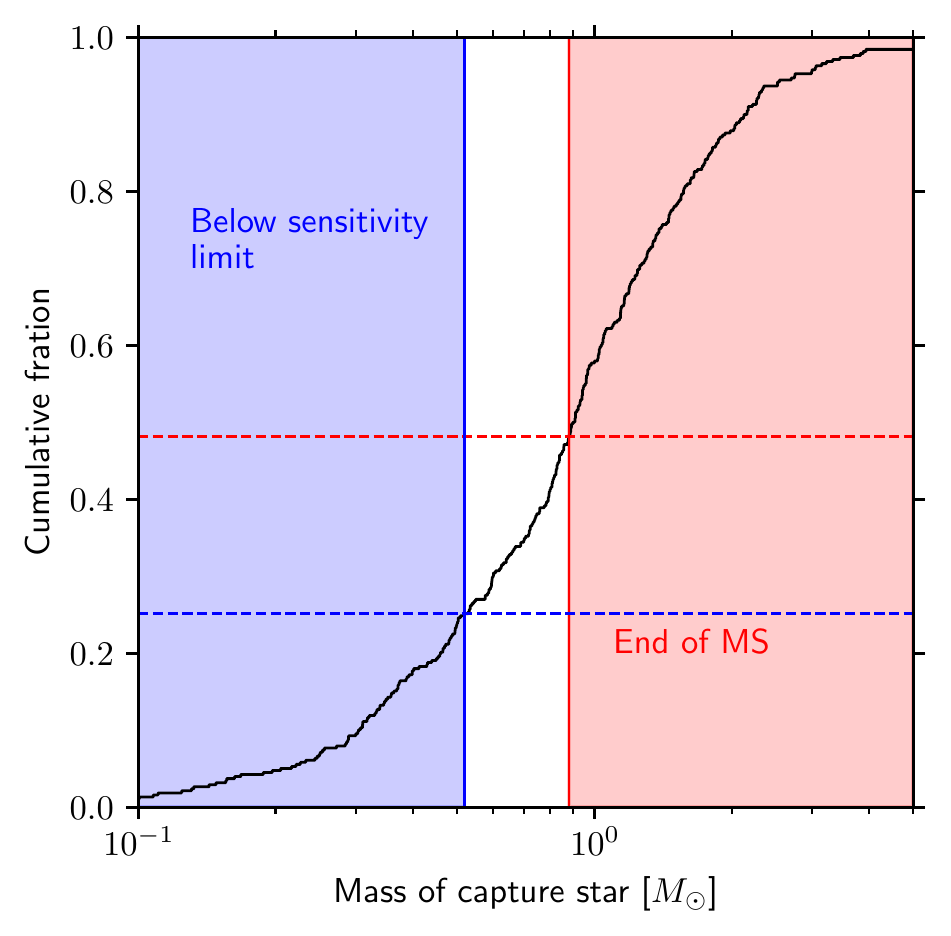}
    \caption{{Cumulative distribution function of the masses of stars that capture a BD over the course of the Monte Carlo simulation of 47 Tuc. The vertical red line is placed at $m_* =0.88\,M_\odot$, for which the visual magnitude in 47 Tuc would be $V\approx 17.1$ according to the models of \citet{Bergbusch92}. This is approximately the mass of the turn off from the main sequence, which are the brightest stars included in the sample of \citetalias{Gil00}. The vertical blue line is at $m_*=0.52\, M_\odot$, which is approximately the cut-off at lower stellar masses adopted by \citetalias[][visual magnitude $V= 21.1$]{Gil00}.}}
    \label{fig:mfunc_capt}
\end{figure}

{In Figure~\ref{fig:mfunc_capt} we show the distribution of stellar masses for stars that capture BDs in the Monte Carlo simulation. We find that many of these captures are for stars of high mass that reach the end of their main sequence lifetime by the present day. Indeed, in the Monte Carlo simulation we find that $210/377$ ($56$ percent) of the captured BDs subsequently undergo a merger with their host due to stellar evolution. The caveat for this finding is that the stellar evolution code employed in \textsc{Mocca} is not necessarily adapted to deal with close BD companions. Whether a closely orbiting BD may survive the end of the main sequence, for example, is therefore unclear. However, we do not explore this concern in this work since \citetalias{Gil00} only included main sequence stars in their survey.}

\begin{figure}
    \centering
    \includegraphics[width=\columnwidth]{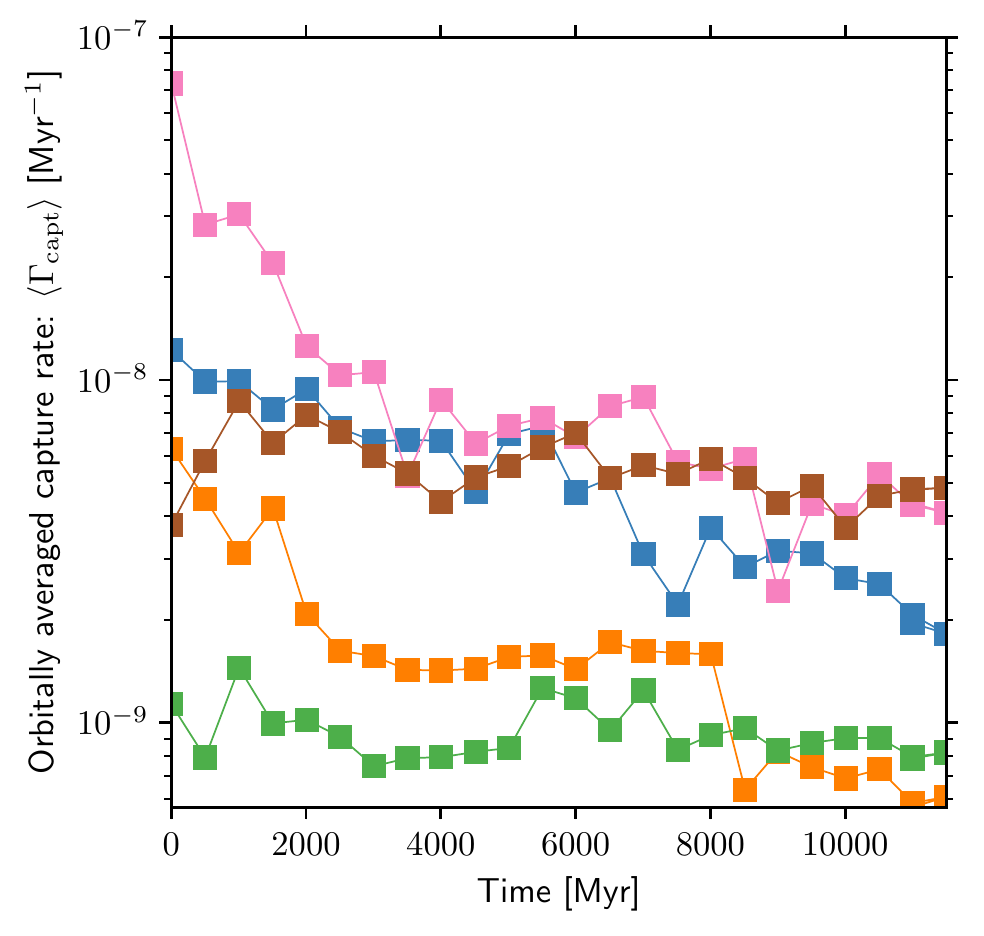}
    \caption{Orbitally averaged BD capture rates for a subset of five stars over the lifetime of 47 Tuc in our Monte Carlo model, {where each star is shown by a different colour line.} The orbital solutions are updated every 500~Myr {and the resultant average capture rate at each update time is denoted by a square marker.}  }
    \label{fig:capt_ints}
\end{figure}

To illustrate how the capture rates vary as a function of stellar mass and final position in the cluster, we compute the orbitally averaged capture rates as outlined in Section~\ref{sec:post-process}. Despite the poor estimate when averaging over the full mass function in Section~\ref{sec:overall_capt} due to mass segregation, we can still estimate per star capture rates. We draw a subset of 1000 stars from the final snapshot, chosen semi-randomly to cover a range of stellar masses and radii within the cluster. We then compute the orbitally averaged capture rate $\langle \Gamma_\mathrm{capt} \rangle$ as defined by equation~\ref{eq:orbavg}, with the orbital solutions and stellar mass updated every $500$~Myr. A subset of five examples are illustrated in Figure~\ref{fig:capt_ints}. We can then integrate these encounter rates over the star lifetimes to give the capture probability (equation~\ref{eq:Penc}). 

\begin{figure*}
    \centering
    \includegraphics[width=0.63\textwidth]{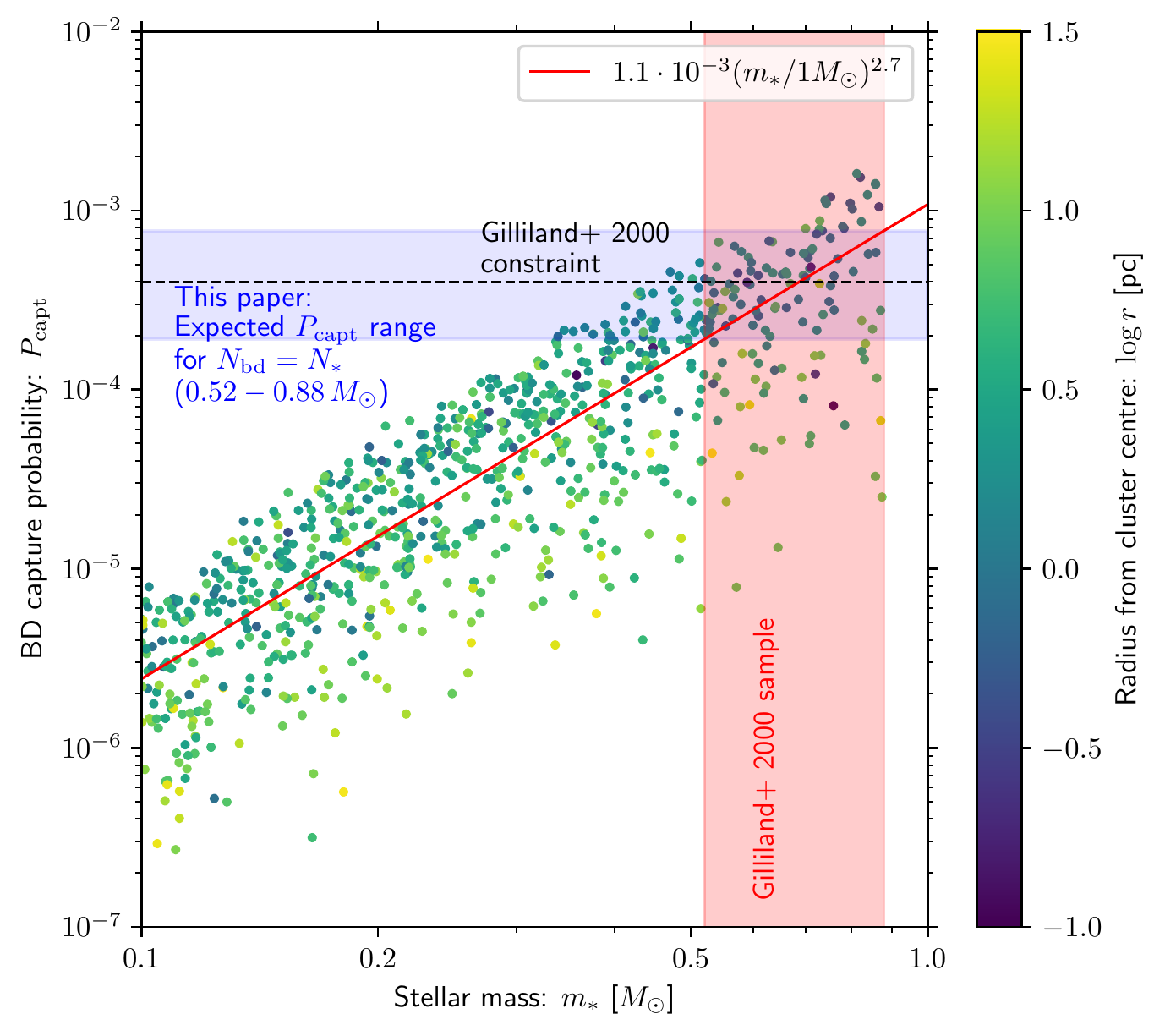}
    \caption{Probability that individual stars have captured a BD over their lifetimes in our dynamical model, as a function of stellar mass. Points are coloured by their radius within the cluster at the end of the simulation. The solid red line is the best power-law fit, while the shaded red region is the mass range of the \citetalias{Gil00} sample. {The corresponding expected range of values of $P_\mathrm{capt}$ for the range of stellar masses surveyed by \citetalias{Gil00} according to the power-law fit (assuming equal numbers of BDs and stars) is shaded in blue. The upper limit on $P_\mathrm{capt}$ inferred by non-detection in that sample is represented by the dashed black line ($P_\mathrm{capt}= 4\times 10^{-4}$).} }
    \label{fig:capture_mfunc}
\end{figure*}

The resulting capture probabilities are shown in Figure~\ref{fig:capture_mfunc}. We find that the probability of capture is a relatively weak function of {final} location, with large scatter. This is the combined influence of the mixing of the stellar population over time and the competing influence of higher velocity dispersion and higher density on the capture rates in the central region (see Section~\ref{sec:theory_rates}).
However, the capture efficiency remains a strong function of stellar mass. {Since masses are spatially segregated, this in turn influences overall spatial dependence of the capture rate such that our assertion that capture probability is not strongly dependent on position is somewhat dependent on sampling in mass-position space. Nonetheless, in an observational context the mass and radial position can be (approximately) measured, such that neither need be marginalized out. }

The exact scaling cannot be computed analytically because it depends non-trivially on the local velocity dispersion evolution and the degree of mass segregation. We therefore fit a function:
\begin{equation}
\label{eq:Pcapt_mst}
    P_\mathrm{capt}(m_*) = A \cdot \left( \frac{m_*}{1\,M_\odot} \right)^{\gamma}
\end{equation} to the capture probabilities for main sequence stars using \texttt{optimize.curve\_fit} in the \textsc{Scipy} software library \citep{Virtanen20}. We obtain $A=1.1\times 10^{-3}$ and $\gamma=2.7$, with the resulting evaluation of equation~\ref{eq:Pcapt_mst} shown as a red line in Figure~\ref{fig:capture_mfunc}. 

\subsection{Comparisons to observational constraints}

\subsubsection{47 Tuc}

{The absence of hot Jupiter detection by \citetalias{Gil00} puts an upper limit on both the fraction of hot Jupiters and tight BD binaries. A close companion with radius $R_\mathrm{comp}$ and separation $a_\mathrm{comp}$ can only be detected by transit if the line-of-sight inclination $i$ satisfies:
\begin{equation}
    \sin i < \frac{R_\mathrm{comp}+R_*}{a_\mathrm{comp}}.
\end{equation} \citetalias{Gil00} adopt $R_\mathrm{comp} = 1.3 R_\mathrm{J}$ and a period of $3.5$~days, corresponding to $a_\mathrm{comp} \approx 9 \, R_\odot$ for a star of mass $m_* \approx 0.7\, M_\odot$. These assumptions lead to the conclusion that $10$ percent of companions should have favourable alignment. This detection efficiency is reduced by a factor $\sim 2$ when aggregating over the stellar radii and time series noise in the sample. However, as noted by \citetalias{Bon03}, the typical tidal capture separations ($a_\mathrm{comp}\approx 4 R_*$) are smaller than the separations assumed by \citetalias{Gil00}. Given this orbit, both the geometric probability and the signal-to-noise are enhanced with respect to the hot Jupiter estimates \citepalias[see Figure 4 of][]{Gil00}. \citetalias{Bon03} estimate a $\sim 20$ percent detection probability for a tidally captured BD. The non-detection of any transit therefore implies an upper limit on the occurrence rate of short period BD binaries of $\lesssim 4\times 10^{-4}$. This limit is represented by the horizontal dashed black line in Figure~\ref{fig:capture_mfunc} and can be compared to the grey region which is the range of $P_\mathrm{capt}$ from equation~\ref{eq:Pcapt_mst} over the range of masses in the \citetalias{Gil00} sample. The theoretical capture probabilities are approximately coincident with the empirical upper limit such that non-detection remains (marginally) consistent with the present empirical constraints.}

\begin{figure}
    \centering
    \includegraphics[width=\columnwidth]{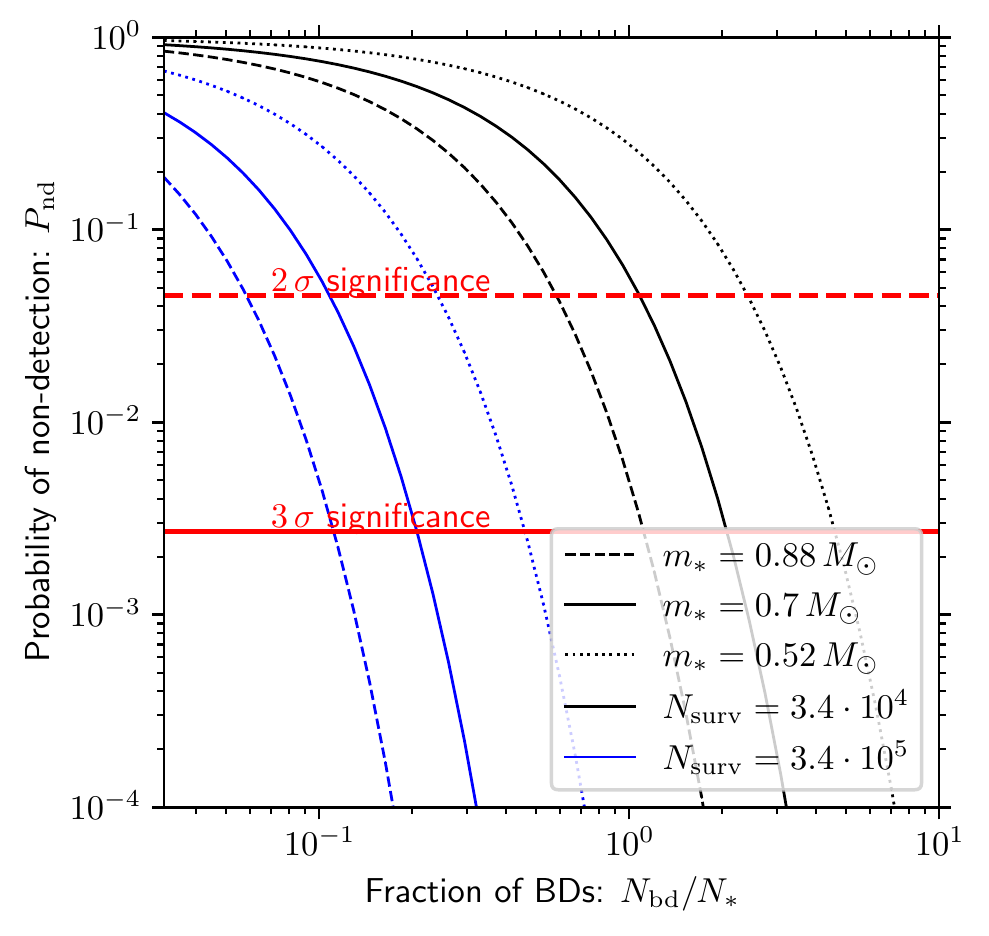}
    \caption{The probability of not detecting any BD in a 47 Tuc-like globular cluster given an initial BD-to-star ratio $f_\mathrm{bd}\equiv N_\mathrm{bd}/N_*$. We have assumed that the total detection efficiency of a transit is $\epsilon_\mathrm{det}=0.2$. We show the present sample size from the \citetalias{Gil00} study in black and an order of magnitude larger sample size in blue. The red lines represent the $2\sigma$ (dashed) and $3\sigma$ (solid) significance for non-detection. The stellar masses are chosen between $m_* = 0.52\,M_\odot$ and $m_*=0.88\, M_\odot$ to bracket the mass range in the study by \citetalias{Gil00}. }
    \label{fig:sigma_nd}
\end{figure}

{More generally, we can write the probability $P_\mathrm{nd}$ of no detection given the number of survey stars $N_\mathrm{surv}$, the capture probability $P_\mathrm{capt}$ and the effective detection efficiency $\epsilon_\mathrm{det}$:
\begin{equation}
    P_\mathrm{nd} = { N_\mathrm{surv}\choose 0 } (1-\epsilon_\mathrm{det} P_\mathrm{capt})^{N_\mathrm{surv}}. 
\end{equation}Adopting $\epsilon_\mathrm{det}=0.2$ and scaling $P_\mathrm{capt}$ in equation~\ref{eq:Penc} by the ratio of BDs to stars $f_\mathrm{bd} \equiv N_\mathrm{bd}/N_*$ yields the results shown in Figure~\ref{fig:sigma_nd}. Adopting $m_*=0.7\, M_\odot$ as a representative star in the sample of \citetalias{Gil00}, we see that $f_\mathrm{bd} \lesssim 1$ to $2\, \sigma$ significance with the current sample of size $N_\mathrm{surv}=3.4\times 10^4$. Increasing the available sample size would (linearly) decrease the maximum $f_\mathrm{bd}$ that is consistent with non-detection. {Future samples should focus on the most massive stars because these offer the best discrimination of the initial BD-to-stellar ratio.  }}

\subsubsection{Other local globular clusters}

\begin{figure}
    \centering
 \includegraphics[width= \columnwidth]{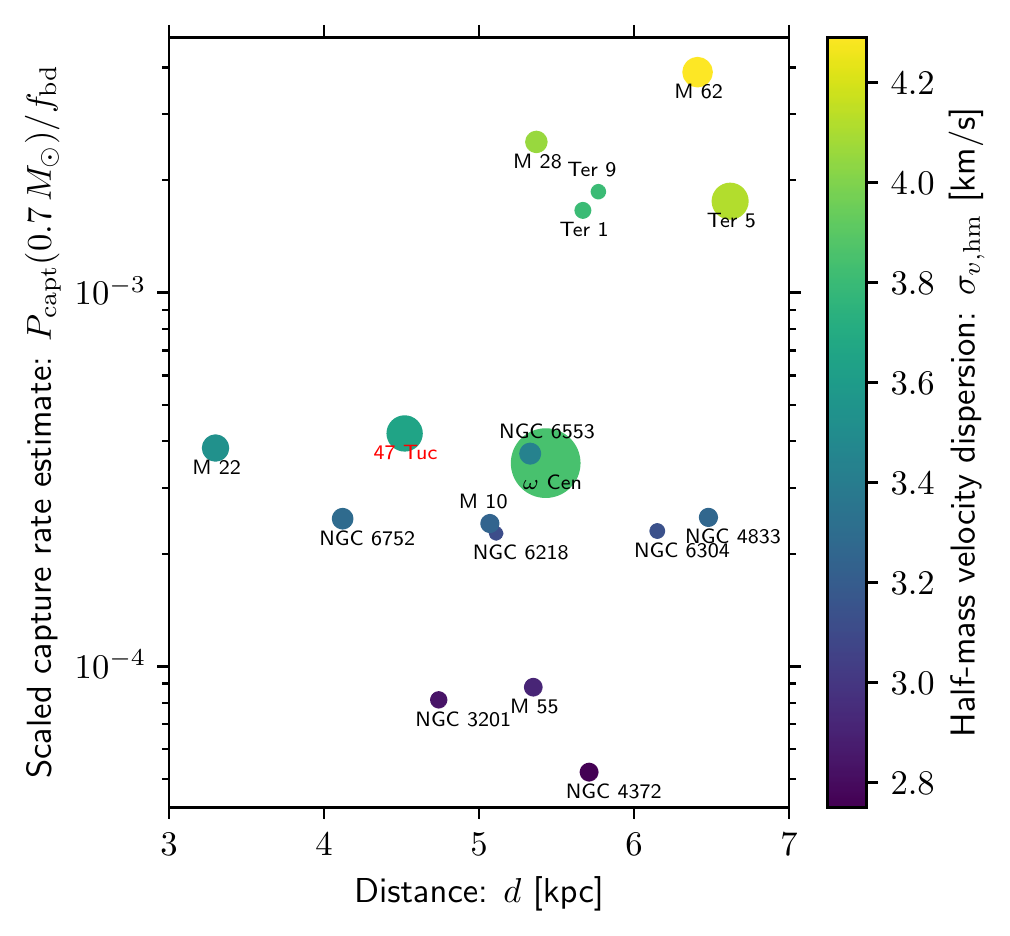}
    \caption{Local globular cluster properties and the estimated capture BD probability for a star of mass $m_*=0.7\, M_\odot$ if the BD-to-star ratio is $f_\mathrm{bd}=1$. Markers are coloured by the one-dimensional half-mass velocity dispersion and {sized proportionally to the total mass.} 47 Tuc (NGC 104) is highlighted in red and all values are summarised in Table~\ref{table:GC_scaled}. }
    \label{fig:GC_plots}
\end{figure}

{To further motivate and guide future observations, we now estimate the equivalent BD capture probabilities in 47 Tuc with respect to other globular clusters. To do this, we adopt parameters from the catalogue of \citet{Hilker20}, who fit N-body simulations to observational data to yield physical property estimates for a large sample, which have been made publicly available.\footnote{\url{https://people.smp.uq.edu.au/HolgerBaumgardt/globular/}} As an order of magnitude estimate, we can then write:
\begin{equation}
    \label{eq:GC_scaled}
    P_\mathrm{capt}^\mathrm{GC} \approx P_\mathrm{capt}^\mathrm{47 Tuc} \frac{n_{*}^{\mathrm{GC}}}{n_{*}^\mathrm{47 Tuc}} \left(\frac{\sigma_{v}^{\mathrm{GC}}}{\sigma_{v}^\mathrm{47 Tuc}} \right)^{-4/3}
\end{equation}where the super-script `GC' denotes an arbitrary globular cluster value and the `47 Tuc' super-scripts are the values for 47 Tuc. We have assumed the power-law $-4/3$ in the velocity scaling, which is the low velocity limit and somewhat shallower than the true scaling. However, without full modelling it is not clear what absolute velocity dispersion for the dominant capture encounters should be adopted. For stars of mass $m_*\approx 0.7\,M_\odot$ and $\sigma_v \lesssim 50$~km~s$^{-1}$ the scaling we adopt is reasonable to make estimates in lieu of full dynamical models.  Note that the exact definitions of both the density $n_*$ and velocity dispersion $\sigma_v$ only matter insofar as we can compare across all clusters, including our reference cluster 47 Tuc. We adopt the half-mass quantities in the database of \citet{Hilker20}, wherein the one dimensional velocity dispersions are quoted.} 

{The relevant physical parameters for a subset of 17 globular clusters within 7~kpc and with masses $>10^5\,M_\odot$ are summarised in Table~\ref{table:GC_scaled} and represented in Figure~\ref{fig:GC_plots}. {We have listed the approximate number of stars $N_{*, 0.52-0.88}$ in the mass range surveyed by \citetalias{Gil00} by integrating the mass function we assume for the 47 Tuc model, truncated above $0.88 \, M_\odot$ in the relevant range.} A number of clusters have significantly higher densities than 47 Tuc, in particular M 28, M 62, Ter1, Ter 5 and Ter 9, which also have considerably higher BD capture probabilities for similar $f_\mathrm{bd}$. {However, these clusters are also relatively close to the galactic centre with low galactic latitude. Their high densities, large distances and possible extinction make these targets challenging for future transit surveys. A number of clusters (e.g. NGC 6656, NGC 6752 and $\omega$ Cen) are relatively nearby and have similar $P_\mathrm{capt}$ to 47 Tuc, thus representing promising targets for future transit surveys. However, overall, 47 Tuc remains possibly the best target given it has a large number of high mass stars that were not surveyed by \citetalias{Gil00} and has the highest estimated $P_\mathrm{capt}$ of the globular clusters that are not in the galactic centre.  }}

\begin{table*}
\centering 
 \begin{tabular}{c c c c c c c c c} 
 \hline
Cluster & $d$ [kpc] & $l$ [$^\circ$] & $b$ [$^\circ$] & $\log \, N_{*,0.52-0.88}$ & $\log n_{*,\mathrm{hm}}$ [pc$^{-3}$] & $\sigma_{v,\mathrm{hm}}$ [km/s] &  $\log [P_\mathrm{capt}(0.7 \, M_\odot)/f_\mathrm{bd}]$  \\
\hline
\rowcolor{Gray}
NGC 104 (47 Tuc) &  $4.52 $& $305.89$ &  $-44.89 $& $5.73$ & $2.35$ & $3.65 $ & $-3.38$ \\
NGC 3201 &  $4.74 $& $277.23$ &  $8.64 $& $4.98$ & $1.49$ & $2.83 $ & $-4.09$ \\
\rowcolor{Gray}
NGC 4372 &  $5.71 $& $300.99$ &  $-9.88 $& $5.07$ & $1.28$ & $2.75 $ & $-4.28$ \\
NGC 4833 &  $6.48 $& $303.60$ &  $-8.02 $& $5.09$ & $2.06$ & $3.26 $ & $-3.60$ \\
\rowcolor{Gray}
NGC 5139 ($\omega$ Cen) &  $5.43 $& $309.10$ &  $14.97 $& $6.34$ & $2.30$ & $3.84 $ & $-3.46$ \\
NGC 6218 &  $5.11 $& $15.72$ &  $26.31 $& $4.80$ & $1.99$ & $3.11 $ & $-3.64$ \\
\rowcolor{Gray}
NGC 6254 (M 10) &  $5.07 $& $15.14$ &  $23.08 $& $5.09$ & $2.04$ & $3.24 $ & $-3.62$ \\
NGC 6266 (M 62) &  $6.41 $& $353.57$ &  $7.32 $& $5.56$ & $3.41$ & $4.29 $ & $-2.41$ \\
\rowcolor{Gray}
NGC 6304 &  $6.15 $& $355.83$ &  $5.38 $& $4.87$ & $2.00$ & $3.13 $ & $-3.64$ \\
Ter 1 &  $5.67 $& $357.56$ &  $0.99 $& $4.95$ & $2.97$ & $3.80 $ & $-2.78$ \\
\rowcolor{Gray}
Ter 5 &  $6.62 $& $3.84$ &  $1.69 $& $5.75$ & $3.04$ & $4.11 $ & $-2.76$ \\
Ter 9 &  $5.77 $& $3.60$ &  $-1.99 $& $4.85$ & $3.02$ & $3.80 $ & $-2.73$ \\
\rowcolor{Gray}
NGC 6553 &  $5.33 $& $5.25$ &  $-3.02 $& $5.23$ & $2.26$ & $3.43 $ & $-3.43$ \\
NGC 6626 (M 28) &  $5.37 $& $7.80$ &  $-5.58 $& $5.25$ & $3.19$ & $4.05 $ & $-2.60$ \\
\rowcolor{Gray}
NGC 6656 (M 22) &  $3.30 $& $9.89$ &  $-7.55 $& $5.45$ & $2.29$ & $3.52 $ & $-3.42$ \\
NGC 6752 &  $4.12 $& $336.49$ &  $-25.63 $& $5.22$ & $2.06$ & $3.28 $ & $-3.60$ \\
\rowcolor{Gray}
NGC 6809 (M 55) &  $5.35 $& $8.79$ &  $-23.27 $& $5.06$ & $1.54$ & $2.91 $ & $-4.06$ \\
\hline
\end{tabular}
\caption{Local globular cluster parameters from the N-body models of \citet{Hilker20}, selected to be closer than $7$~kpc and more massive than $10^5\, M_\odot$. {The number of stars $N_{*, 0.52-0.88}$ is estimated by dividing the total mass by $0.5\,M_\odot$ then multiplying by $0.297$, the approximate fraction of stars with masses $0.52 - 0.88\,M_\odot$.} The stellar number density, $n_*$, and (one dimensional) velocity dispersion, $\sigma_v$ are taken inside the half-mass radius. The last column is obtained by scaling the results for 47 Tuc using equation~\ref{eq:GC_scaled}. }
\label{table:GC_scaled}
\end{table*}

\subsection{Caveats for capture rates}
\label{sec:caveats}

{We have explored BD capture rates in detail and suggested that sufficiently large transit surveys can put upper limits on BD formation rates. However, it is possible that the present day short period companion rates are influenced by other physical mechanisms. Factors that may alter the rates of short period BD companions include (although not necessarily limited to): }
\begin{itemize}
    \item \textit{Primordial mass segregation}: We have demonstrated that once a cluster become mass segregated, BD tidal capture becomes inefficient. If a population is primordially segregated, this would similarly reduce the capture efficiency.
    \item \textit{Time-scale for circularisation:} In Paper II we explore the time-scale on which a migrating planet may undergo a dynamical perturbation while circularising. Following \citetalias{Bon03}, we have assumed that this time-scale is short for a tidally captured BD \citep{Mardling96}. However, if this is not the case then perturbations after the initial tidal encounter may curtail tidal circularisation and therefore prevent the formation of the {tight BD-star binary that can be detected through transit}. 
    \item \textit{Tidal inspiral of BDs:} {Evidence for the correlation of hot Jupiter occurrence with cold stellar kinematics may originate from the inspiral of close companions onto the central star on Gyr timescales \citep[][]{Hamer19}. If close sub-stellar companions do inspiral on these time-scales, then a similar process {may operate on} tidally captured BDs. However, hot Jupiters appear to be retained in the dense cluster M67 \citep[][]{Brucalassi16} which has an age of $\sim 4.5$~Gyr, such that this would require a relatively narrow range of inspiral timescales \citep[see also discussion in Section 4.3 of][]{Winter21}.}
    \item \textit{Evacuation of BDs:} Apart from mass segregation due to two-body relaxation, low mass stars and BDs can be further evacuated from the central regions of the globular cluster by alternative heating mechanisms. For example, black hole subsystems may induce dynamically heating and eject low mass objects such as BDs to the cluster halo \citep{Breen13,Giersz19}. However, this process occurs on a time-scale longer than the half-mass relaxation time-scale ($\sim 3$~Gyr for 47 Tuc). The time-scale on which the majority of BD captures occurs in our models is $\lesssim 2$~Gyr, while segregation on longer time-scales may not strongly influence capture rates. Similarly, heating due to tidal shocks during to passages through the galactic plane may operate time-scales comparable to two-body relaxation \citep{Gnedin99}. If these mechanisms significantly reduce the stellar density after $\sim 12$~Gyr, this would also suggest a moderately higher initial stellar density required to reproduce the present day density profile. These considerations may therefore increase the initial capture rate and subsequently reduce it due to enhanced mass segregation. We do not explore these possibilities quantitatively in this work. 
\end{itemize}

{An absence of close sub-stellar companions would therefore suggest either that BD formation is suppressed or that one of the above processes (or unconsidered alternative) is operating. In any case, non-detection in an increased sample of stars would require explanation and future survey campaigns are therefore merited.}

\section{Conclusions}
\label{sec:conclusions}

In this work, we have explored the apparent absence of close-in sub-stellar companions in the globular cluster 47 Tuc from a theoretical perspective. We applied a Monte Carlo model using the \textsc{Mocca} code \citep{Hyp13, Giersz13} for the dynamical evolution of the globular cluster. Using this model, we compute the rates of tidal BD capture over its lifetime.

Our results indicate lower capture efficiency than previous estimates \citep{Bon03}. The reasons for this are subtle, but fundamentally originate from the rapid decrease of the tidal capture cross section with decreasing stellar mass. This is particularly true for environments with velocity dispersions as high as globular clusters. Once mass segregation operates, BDs and low mass stars are preferentially found in the same spatial location. Therefore the global tidal capture efficiency drops precipitously, such that the current constraints {cannot rule out that the frequency of BDs in the IMF is as high in 47 Tuc as in the galactic field.}

These considerations also lead to a steep scaling of the capture probability with stellar mass. For initial number of BDs $N_\mathrm{bd}$ and stars $N_*$, those stars that have not reached the end of their main sequence have a lifetime capture probability:
\begin{equation}
    P_\mathrm{capt} = 1.1 \times 10^{-3} \frac{N_\mathrm{bd}}{N_*}\cdot \left(\frac{m_*}{1\, M_\odot} \right)^{2.7}.
\end{equation}The large exponent means that any constraints on the initial BD ratio are strongly dependent on the mass function of stars that are surveyed for close companions. For the typical masses of the stars surveyed by \citet[][$0.52 \,M_\odot \lesssim m_* \lesssim 0.88\, M_\odot$]{Gil00} and equal numbers of BDs and stars, this yields capture probabilities that are comparable to the upper limit constraint on close sub-stellar companions ($P_\mathrm{capt}\lesssim 4\cdot 10^{-4}$).


Finally we conclude that, while the current constraints on the frequency of close sub-stellar companions {cannot rule out that the incidence of BDs in 47 Tuc is as high as it is in the field,} stronger constraints can be obtained by surveying a larger number of relatively high mass stars. {Such an exercise may also be achieved aggregating across several globular clusters.} We therefore estimate the capture rates in local globular clusters for a similar mass range of stars to those surveyed in 47 Tuc. The estimated capture rates are summarised in Table~\ref{table:GC_scaled}. {We suggest that 47 Tuc remains among the most promising targets for follow up, with a convenient location and a large number of relatively high mass stars that have not yet been monitored for short period sub-stellar companions. A number of other globular clusters, such as $\omega$ Cen, may also represent feasible targets for transit surveys. Our results offer motivation and interpretation for future transit surveys of globular clusters.}

\section*{Acknowledgements}

We thank the anonymous referee for their careful reading that helped clarify the manuscript and Abbas Askar for his helpful comments on dynamical models for 47 Tucanae. AJW acknowledges funding from an Alexander von Humboldt Stiftung Postdoctoral Research Fellowship. GR acknowledges support from the Netherlands Organisation for Scientific Research (NWO, program number 016.Veni.192.233) and from an STFC Ernest Rutherford Fellowship (grant number ST/T003855/1). This project has received funding from the European Research Council (ERC) under the European Union’s Horizon 2020 research and innovation programme (grant agreement No 681601) and been supported by the DISCSIM project, grant agreement 341137 funded by the ERC under ERC-2013-ADG. 

\section*{Data availability}

All data in this article is available from the corresponding author upon reasonable request.




\bibliographystyle{mnras}
\bibliography{bdbib} 


\appendix

\section{Variable mass function}
\label{app:massfunc}

\begin{figure}
    \centering
   \subfloat[\label{subfig:Gammahat}Overall per star capture rate and initial mass function]{\includegraphics[width= 0.47\textwidth]{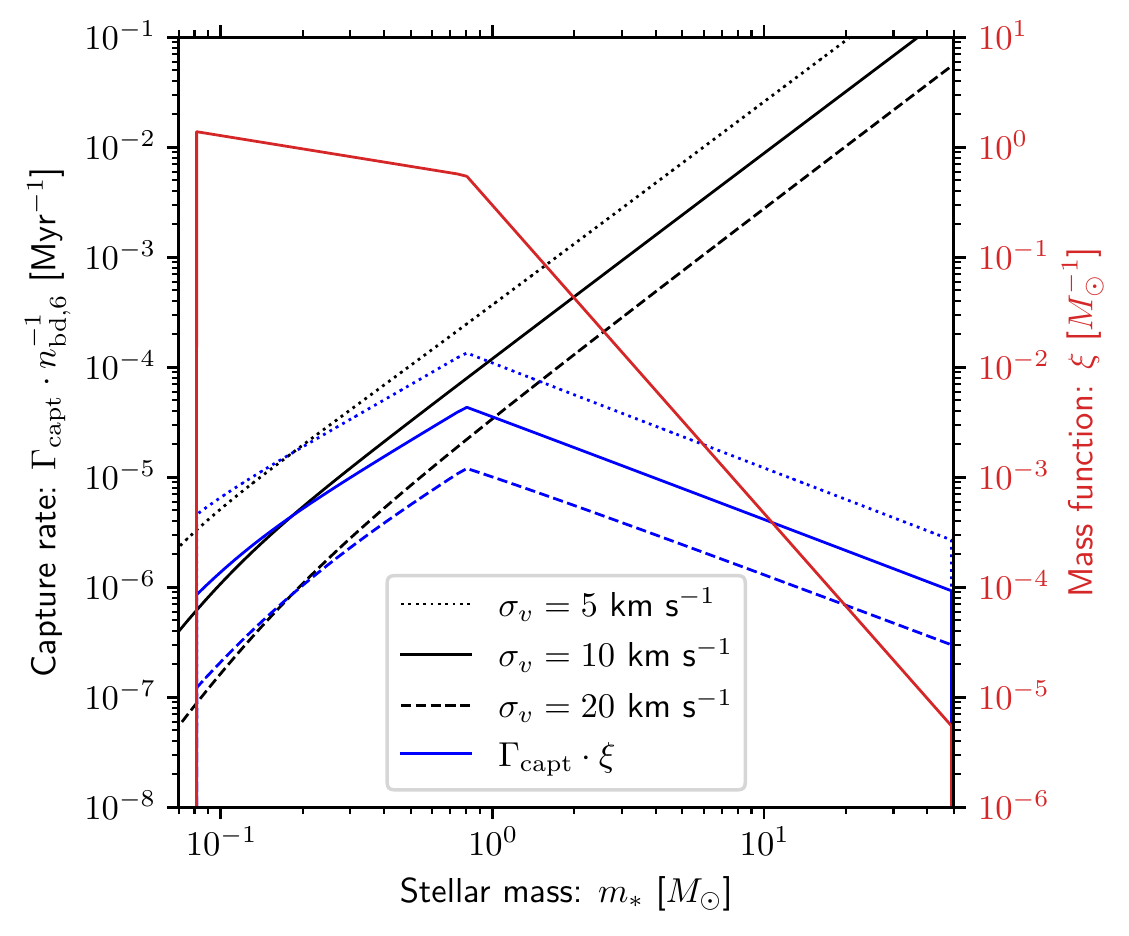}}\\
        \subfloat[\label{subfig:Gammahat_mfv}Relative capture rate with maximum mass]{\includegraphics[width= 0.4\textwidth]{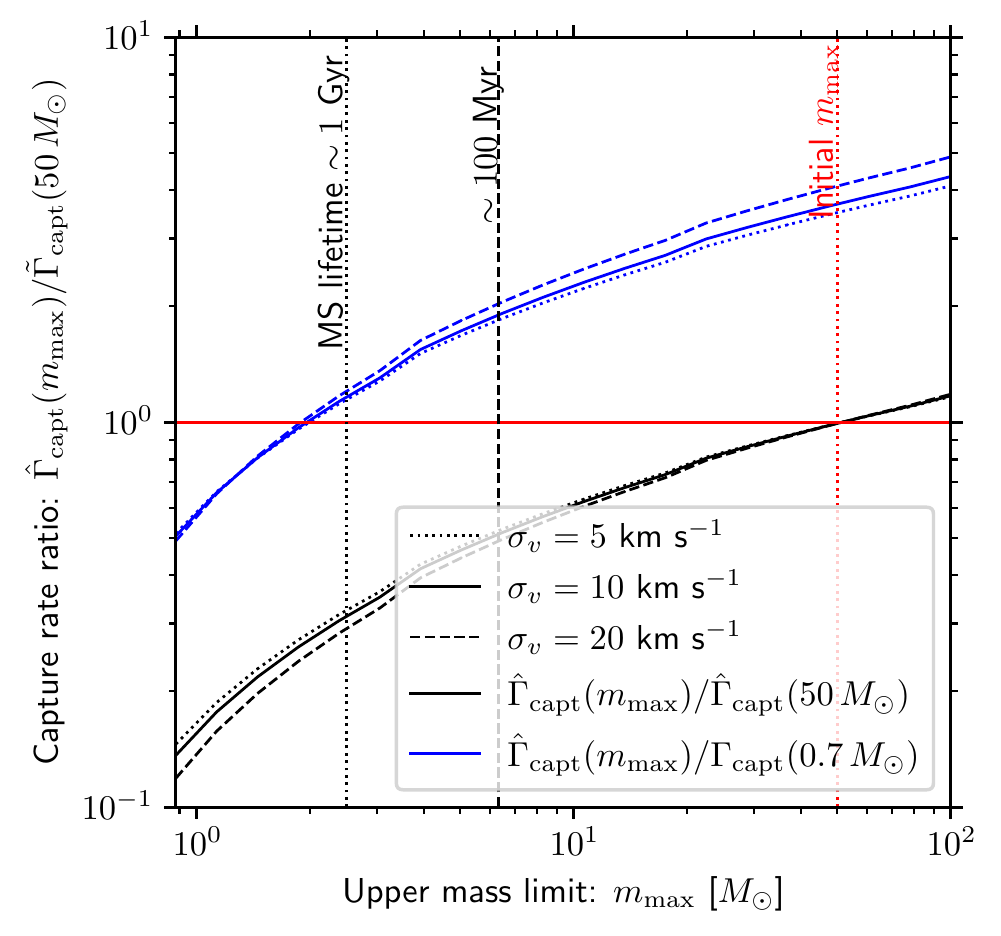}}\\
    \caption{Relative encounter rate with varying stellar mass (function). The black lines in Figure~\ref{subfig:Gammahat} shows the capture rate (integral of equation~\ref{eq:Gamma_capt}) as a function of stellar mass and corresponding radius given by equation~\ref{eq:approx_captrate}, while the red line shows the initial mass function we assume for the Monte Carlo model. The blue lines show the product of the mass function and the capture rate. In Figure~\ref{subfig:Gammahat_mfv} we show the variation of the overall capture rate $\hat\Gamma_\mathrm{capt}$ integrated across all masses truncated above the upper mass limit $m_\mathrm{max}$. The black lines are normalised by the equivalent capture rate with the initial mass function ($m_\mathrm{max} = 50\,M_\odot$) and the blue lines are normalised by $\Gamma_\mathrm{capt}$ for a single stellar mass ($m_*=0.7\, M_\odot$). The vertical lines represent the initial most massive star (red dotted), the most massive star after $100$~Myr (black dashed) and the most massive star after $1$~Gyr (black dotted). }
    \label{fig:Gammahatcapt}
\end{figure}

To quantify the degree to which the mass function determines the capture rates, we define the average capture rate:
\begin{equation}
\label{eq:Gammahat}
    \hat{\Gamma}_\mathrm{capt}(\Theta) =  \int_{m_{\mathrm{min}}}^{m_\mathrm{max}}  \, \mathrm{d} m_* \, \xi(m_*)  \Gamma_\mathrm{capt} (m_*; \Theta),
\end{equation} where $\Theta$ is an arbitrary variable representing the remaining physical quantities that determine the local BD capture rate for a given star. Here $\xi(m_*)$ is the stellar mass function. The initial mass function we adopt in our Monte Carlo model is a broken power-law with parameters described in Table~\ref{table:dyn_mods}, between the minimum mass $m_{\rm{min}}=0.08\,M_\odot$ and maximum mass $m_{\rm{max}} = 50\, M_\odot$. For the purposes of estimating the encounter rates, we adopt a simple mass-radius relation:
\begin{equation}
\label{eq:approx_mr}
    R_* = \left(\frac{m_*}{1\,M_\odot}\right)^{0.8} \,R_\odot,
\end{equation}additionally fixing $q=0.08 M_\odot/ m_*$ and $R_\mathrm{bd}=0.1 \, R_\odot$ to leave only extrinsic properties ($n_\mathrm{bd}$ and $\sigma_v$) in $\Theta$. We show the product of the per star capture rate and the initial stellar mass function (i.e. the integrand of equation~\ref{eq:Gammahat}) in Figure~\ref{subfig:Gammahat}.  

To approximate the decrease in the number of high mass stars over time as they reach the end of their lifetimes, we truncate the mass function above variable mass $m_\mathrm{max}$. We then compare the $\hat{\Gamma}_\mathrm{capt}$ we obtain to that obtained when averaging over our initial mass-function. The results are shown in Figure~\ref{subfig:Gammahat_mfv}, showing an approximately order of magnitude decline in overall capture efficiency as the massive stars are removed from the mass function. This means that an accurate mass function over time is required to estimate the total number of encounters. This would strictly require resolving the functional form of the mass function, both locally and over the lifetime of 47 Tuc. However, we also show that adopting $\Gamma_\mathrm{capt} (m_* = 0.7\,M_\odot)$ (blue line in Figure~\ref{subfig:Gammahat_mfv}) produces a reasonable estimate of the overall capture rate after a short time ($\sim 100$~Myr), once the most massive stars have reached the end of their lifetimes. We therefore adopt this approximation for the overall capture rate when we validate our Monte Carlo results in Section~\ref{sec:bd_capt}. 

\bsp	
\label{lastpage}
\end{document}